\documentclass[10pt,a4paper]{article}

\usepackage{amsmath,amssymb,amsfonts,graphicx,dcolumn}
\usepackage{epstopdf}
\usepackage{color}
\usepackage[utf8]{inputenc}
\usepackage{amsthm}
\usepackage{fontenc}
\usepackage{tikz}
\usepackage{hyperref}
\RequirePackage{slashed}
\usepackage{cancel}
\usepackage[normalem]{ulem}

\newcommand{\se}{\sigma_{eff}}
\newcommand{\mse}{\overline{ \sigma}_{eff}}
\usepackage{comment}
\usepackage{color, colortbl}
\definecolor{Gray}{gray}{0.9}
\definecolor{LightCyan}{rgb}{0.88,1,1}
\usepackage{makecell}

\usepackage{caption}
\usepackage[titletoc]{appendix}

\usepackage[a4paper,top=2cm,bottom=2cm,left=2cm,right=2cm]{geometry}

\title{Double parton correlations in mesons within AdS/QCD soft-wall 
models: a first comparison with lattice data}

\author{Matteo Rinaldi \\ \\ Universit\`a degli studi di 
Perugia, 
Dipartimento di Fisica 
e Geologia.  \\ INFN Sezione di Perugia. Via A. Pascoli, 06123, 
Perugia,  Italia.}

\date{}

\begin{document}

\maketitle

\begin{abstract}

Double parton distribution functions (dPDFs),  
entering the double parton scattering (DPS) cross section, are  unknown
fundamental 
quantities encoding new interesting properties of hadrons. Here, the 
pion dPDFs are investigated within different holographic QCD 
quark models in order to access their basic features. 
Results of the calculation,s obtained within the AdS/QCD soft-wall 
approach, 
 have been compared with predictions of 
lattice QCD evaluations of the pion two-current correlation functions. 
The present 
analysis confirms that double parton correlations, affecting dPDFs, are 
very important and not direct accessible from generalised parton 
distribution functions and electromagnetic form factors.
The comparison between lattice data and quark 
model calculations
unveils the relevance
of the 
contributions of high partonic Fock states in the pion.
Nevertheless, by using a complete general procedure, results of lattice 
QCD have been used, for the first time, to estimate the mean value of 
the so called $\se$, 
a relevant 
experimental observable  for DPS processes.
 In addition, the results of the 
first 
calculations of the $\rho$ meson dPDFs are  discussed in 
order to 
make predictions.    

\end{abstract}
 
 \newpage
\tableofcontents
\newpage
\section{\label{sec:intro}Introduction}

In the last few years,  great attention has been devoted to  
theoretical and experimental studies of multiple parton interactions 
(MPI), due to the large demand of detailed  description of 
hadronic final states required at the LHC \cite{Paver:1982yp,1987}. The 
inclusion of MPI in experimental analyses is fundamental for the 
research 
of New Physics, being MPI a source of background.
The simplest case of MPI is the double parton 
scattering (DPS) \cite{Goebel:1979mi,Mekhfi:1983az}, where two partons 
of an hadron simultaneously interact with other two partons of the 
other 
colliding hadron. As discussed in a recent review \cite{scokas}, the 
measurements of DPS processes are mandatory to access unknown double 
parton correlations (DPCs) in the proton. Moreover, the 
DPS cross section depends on a new quantity called double parton 
distribution functions (dPFDs) which encode the probability of finding 
two partons, with given flavors,  longitudinal momentum fractions 
($x_1,x_2$) and relative transverse distance $d_\perp$ 
\cite{Paver:1982yp,Diehl1,Calucci:1999yz}.
 If measured, dPDFs would therefore represent 
a novel tool to access the three-dimensional hadron structure 
\cite{noij3,rapid}. In fact,
dPDFs provide  new fundamental information, complementary to those 
obtained by using 
generalised parton distribution functions (GPDs) \cite{burk}.
However, for the moment being, no  data for the proton dPDFs have been 
so far 
collected. Furthermore,  dPDFs  are non perturbative objects in QCD
not  directly accessible from the theory. It is therefore useful to
estimate them at low momentum scales ($\sim \Lambda_{QCD}$),
for example by  using quark models~ 
\cite{Mel_19,Mel_23,noi2,noi1}. 
In addition to several general analyses on dPDFs~
\cite{Diehl1,noij3,gaunt1,gaunt2,Cotogno:2020iio,cotogno2},   a 
lattice QCD investigation on two-current correlations in the pion has 
been published very recently~ \cite{lattice}. In the present analysis, 
we 
take advantage of the lattice data, to test quark model predictions for 
the pion dPDFs.
The  
calculations of the latter have been shown for the first time in Ref. 
\cite{noipion} 
and then  within the Nambu-Jona-Lasinio (NJL) model~ 
\cite{njl,njl2,njl3}.
In particular, by following 
the line of Ref. \cite{noipion}, we consider here  AdS/QCD soft-wall 
quark models. Let us mention that the mean value of $\se$, 
sensitive to dPDFs, has been already calculated within an holographic 
QCD model 
for the 
proton target~ \cite{Traini:2016jru}. 
The models here used are  inspired by the so called AdS/CFT 
correspondence \cite{maldacena,Witten:1998qj},  which relates a 
supersymmetric conformal field theory with a classical graviational one 
 in an 
anti-de-Sitter space. In the so called
bottom-up approach, one implements fundamental properties of QCD by 
generating a 
theory 
in which 
conformal symmetry is asymptotically   { restored}~
\cite{Brodsky:2014yha,Polchinski:2000uf,Polchinski:2001tt,Brodsky:2003px,Erlich:2005qh}.
Let us mention that this 
approach has been successfully applied
to access non perturbative features  of QCD, for example
 the description of the 
spectrum of glueballs, 
hadrons,  form factors { (ffs)} and different kind of parton 
distribution functions (PDFs)~
\cite{Br1,Br2,universal,rho,rinaldiGPD,
glue1,glue2,glue3,traini,pasquini1,spin,Chakrabarti:2013gra}.
In the present investigation we discuss the calculations of the pion 
dPDFs  and their first moments within the AdS/QCD approach. Comparisons 
with lattice outcomes will test the predictive power of these models 
and 
provide new fundamental constraints for their future improvements. 
In the last part of the present investigation, predictions for the
$\rho$ dPDFs will be shown for the first time.
The 
paper is organised as follows.

In Sect. \ref{II} the  formalism to describe dPDFs and related 
quantities within the Light-Front (LF) approach is shown.
In Sect. \ref{III} a brief recapitulation of the main lattice 
evaluation of 
moments of  dPDFs \cite{lattice} will be presented. In Sect. \ref{IV} 
details on the adopted AdS/QCD  models  will be discussed. In 
Sect. \ref{V} numerical calculations of dPDFs and related quantities 
will be shown, also including the comparisons with lattice data.
In Sect. \ref{VI} the first study of the $\rho$ dPDFs is presented.

\section{\label{II} Meson Double PDF within the Light-Front approach}
In this formal section, the main strategy to obtain a suitable 
expression of the mesonic dPDFs,  for quark model calculations, 
will be presented. In particular, the essential steps of this procedure 
have been previously developed in Refs. \cite{noipion,kase} and they 
are 
here 
summarised. In particular we consider the Light-Front (LF) approach 
\cite{58Br,keister} together with the LF wave function 
representation of the hadronic state~\cite{ov1,ov2}. 
In this scenario, the meson (M) state $|M,P \rangle$, with momentum 
$P^\mu$, can be decomposed in a coherent sum over partonic Fock states. 
The relative contribution of a given Fock state to the meson is 
encoded 
in the so 
called LF wave function (w.f.) $\psi$. The latter contains all 
non-perturbative information on the meson structure. Of course, 
the LF w.f. cannot be evaluated from first principles, i.e. the QCD. In 
this scenario, constituent quark models represent suitable tools to 
evaluate the w.f. and then to 
explore basic non perturbative features of different kind of 
distributions, such as parton 
distribution functions, form factors and dPDFs. Indeed, all these 
quantities can be 
described in terms of the LF wave function.
In the present analysis we focus our attention on dPDFs.
As already mentioned,
these quantities 
encodes novel  information on the hadron structure which 
cannot be obtained through one-body functions such as generalised 
parton distributions  and transverse momentum dependent PDFs 
(TMDs).
Since the main purpose of this investigation is the comparison between 
quark model calculations with those obtained within the lattice 
framework 
\cite{lattice}, here we
 consider
 the unpolarized
dPDFs which depend on the Dirac 
matrix  $\gamma^\mu$. 
The double PDFs can be formally defined through  
a
light-cone correlator \cite{Diehl1}: 

\begin{align}
\label{f2c1}
 & F_{q_1 q_2}(x_1,x_2, {\bf k_\perp} ) = {P^+ \over 4 } \int d^2 {\bf 
y_\perp}
  e^{-i 
{\bf 
y_\perp} \cdot {\bf k_\perp} }\int dy^-
 \int dz^-_1dz_2^-
 \\
\nonumber
&\times
 {e^{-i x_1P^+ z^-_1-i 
x_2P^+ z^-_2} \over (2\pi)^2 }
  \langle M, {\bf 0}| \mathcal{O}_{q_1}(0,z_1) 
\mathcal{O}_{q_2}(y,z_2)|M, {\bf 
 {\bf 0}} \rangle \Big|_{y^+ =z^+_1=z_2^+=0}^{{\bf z_{1\perp}= 
z_{2\perp}  }=0  
}~,
\end{align}
where, for generic 4-vectors $y$ and $z$, the operator 
$\mathcal{O}_{q}(y,z)$ 
{ for the quark of flavor $q$} 
reads:
 
 \begin{align}
 \label{op}
  \mathcal{O}_{q}(y,z) = 
  \bar q\left( y- {1 \over 2} z  \right) \gamma^+ 
 q 
\left(  
 y+ {1 \over 2}z \right)~,
 \end{align}
\\  and $q(z)$ is the LF quark field operator.
In order to find a suitable expression of the dPDF, we consider the 
Fock decomposition of the mesonic state \cite{ov1,ov2} and keep  only
the  $|q \bar q \rangle$
contribution \cite{noipion}. In fact,  for the moment 
being,  an 
explicit expression for the 
LF wave function of, e.g., the  $|q \bar q q \bar q \rangle$ 
state, 
is not available. Therefore, the meson state reads:

\begin{align}
\label{state}
\nonumber
 |M, {\bf P_\perp }  \rangle &= \sum_{h, \bar h} \int 
\dfrac{dx_1~dx_2}{\sqrt{x_1 x_2}} \dfrac{d^2 {\bf k_{1\perp}} d^2 \bf 
k_{2 
\perp} }{2(2 \pi)^3}\delta^{(2)}({\bf k_{1 \perp}+ k_{2\perp}} )
|x_1, {\bf k_{1 \perp}}+ x_1 {\bf P_\perp},h \rangle |x_2, 
{\bf k_{2 
\perp}}+ x_2 {\bf P_\perp},\bar h \rangle
\\
&\times ~\delta(1-x_1-x_2) \psi^M_{h, 
\bar h} (x_1,x_2,{\bf k_{1 \perp},k_{2\perp}}  )~.
\end{align}
 \\
 { Here,}
$h$ and $\bar h$  represent  the
parton helicities, $x_i = k_i^+/P^+$ and ${\bf k_{i 
\perp}}$ the 
quark longitudinal momentum fraction and its transverse momentum, 
respectively, and
$P^\mu$ is  the meson 4-momentum. The light cone components of a 
generic 
4-vector are defined 
 by
$l^\pm = l^0 \pm l^3$. 
In Eq. (\ref{state}),  $ \psi^M_{h, 
\bar h} (x_1,x_2,{\bf k_{1 \perp},k_{2\perp}}  )$ is the  LF meson 
wave-function, whose normalisation  { is } chosen to be
\begin{align}
\label{norm}
  {1 \over 2} \sum_{h,\bar h} \int dx_1 dx_2  {d^2 {\bf k_{1\perp} }  
 d^2 
{\bf k_{2\perp} }\over 16 \pi^3}  \delta(1-x_1-x_2)
 \delta^{(2)}({\bf k_{1 \perp} {\bf +} k_{2\perp}}) 
|\psi^M_{h, 
\bar h} (x_1,x_2,{\bf k_{1 \perp},k_{2\perp}}  )  |^2 =1~.
\end{align}
  { The w.f.  $ \psi^M_{h, 
\bar h} (x_1,x_2,{\bf k_{1 \perp},k_{2\perp}}  ) $   } determines the 
structure of the state.  
The direct expression of the dPDF in terms of the above quantity can be 
obtained by following the procedure developed in 
 Refs. 
\cite{noi1,noipion,kase}. In the Appendix A, details on 
the convention for 
 the quark-antiquark  
field 
operator and  anticommutation relations, between 
creation-annihilation operators
 \cite{Br1}, are shown.
Finally, the 
meson dPDF reads:

\begin{align}
\nonumber
 F_{q_1 \bar q_2}(x_1,x_2, {\bf k_\perp} ) &=
 {1 \over 2} \sum_{h,\bar 
h} \int 
{d^2{\bf 
k_{1\perp}} 
\over 2 (2 
\pi)^3}  \psi^M_{h, 
\bar h} (x_1,x_2,{\bf k_{1 \perp}},-{\bf k_{1 \perp}}   )
\psi_{h, 
\bar h}^{*M} (x_1,x_2,{\bf k_{1 \perp}+k_\perp},-{\bf k_{1 
\perp}-k_\perp}  )
 \delta(1-x_1-x_2)
\\
&= f_2^M(x_1,{\bf k_\perp} )\delta(1-x_1-x_2).  
\label{fdiv}
\end{align}
{ In the above expression, $q_1$ and $\bar q_2$ are the flavors of the 
constituent quarks. } Due to momentum conservation,  $x_2 = 1-x_1$ 
and $ {\bf k_{2 \perp}}=-{\bf k_{1 \perp} }
  $; thus  we define  $ \psi_{h, 
\bar h} (x_1,{\bf k_{1 \perp}}  )= \psi^\pi_{h, 
\bar h} (x_1,1-x_1,{\bf k_{1 \perp}},{\bf -k_{1\perp} }  )$ for brevity.

Since as already mentioned, the comparison with lattice data is 
fundamental
in the present investigation, we are mainly interested in moments of 
dPDFs, i.e. the integrals over $x_1$ and $x_2$ of Eq. (\ref{fdiv}). 
Thus 
$f_2^M(x_1, {\bf k_\perp} )$ in Eq. (\ref{fdiv}), 
is the quantity that will be calculated within constituent quark models:

\begin{align}
\label{f2m}
 f_2^M(x, {\bf k_\perp} ) &= 
 \int_0^1 dx_2~ F_{q_1 \bar q_2}(x,x_2,{\bf 
k_\perp})
=
 {1 \over 2} \sum_{h,\bar h} 
 \int {d^2{\bf k_{1\perp}} 
\over 2 (2 
\pi)^3}  \psi^M_{h, 
\bar h} (x,{\bf k_{1 \perp}}  ) \psi_{h, 
\bar h}^{*M} (x,{\bf k_{1 \perp}+k_\perp}  ).  
\end{align}
As shown in Refs. \cite{noi1,gaunt1,gaunt2}, dPDFs evaluated at
 ${\bf k_\perp}=0 $ are related to
the PDF.
Here and in the following, the meson PDFs are specified by the 
subscript ``1'' , i.e. $f_1^M(x)$.
As one might notice, if only a two-body Fock state
is considered in  Eq. (\ref{state}),  the dPDF would be essentially an 
unintegrated PDF.
In the proton case, where  the $|qqq\rangle$ state is the dominant 
one, the above feature is not valid.
In this analysis we make use of different quark models to 
identify  general non perturbative features of  dPDFs. Therefore, the 
following ratio is studied \cite{noi2,noi1}  to    emphasise  the role 
of 
correlations 
between the $x$ and ${\bf 
k_\perp}$ dependence:

\begin{align}
\label{rk}
 r_k(x, k_\perp)   ={ f_2^M(x,k_\perp)  \over 
f_2^M(0.4,k_\perp)  }~;
\end{align}
in   fact, if a factorised ansatz  for dPDFs  were 
valid,
 for example 
$f_2^M(x,k_\perp) \sim f_{2,x}(x) f_{2, k_\perp}(k_\perp)$, then the 
ratio
 $r_k(x, k_\perp)$   would  not depend on 
$k_\perp$.
For details on the calculations of this quantity, in the proton case, 
see Refs. 
\cite{noi2,noi1,noij3,melo}.
 Let us remind that this kind 
of 
 ansatz is often 
used 
in experimental analyses.

{ In closing this section, we note that the dPDFs 
depend on two momentum scales. 
{ Therefore, in order to make useful predictions, the perturbative 
QCD evolution of dPDFs should be properly included in the analyses. 
Moreover, as shown in several papers, see e.g. Refs. \cite{noi1,noij2}, 
the 
evolution procedure can reduce the impact of $x_1-x_2$ correlations. 
However,
 since the pQCD evolution equations of dPDFs do not 
involve the $k_\perp$ dependence, correlations between $x$ and 
$k_\perp$ can be relevant also at high energy scales.
Such a conclusion has been discussed in Ref. \cite{noipion} for the 
pion, and   in Refs. \cite{noij3,melo}  for the proton.}
Furthermore, since for the moment being we are mainly interested in the 
first 
moment of 
dPDFs, we 
take 
both scales  equal to the hadronic one. For evolution effects in the 
pion 
dPDFs 
see Ref. \cite{noipion}.}

\subsection{Moments of dPDFs}
As already mentioned,
in the present study we are mainly interested on the first
moment of the pion dPDF. Results of the calculations of this quantity 
will be  
compared to that 
obtained within the lattice 
 \cite{lattice}. 
 The physical 
interpretation of the first moment of dPDFs is here discussed. 
As shown in Refs. \cite{noij3,noiplb,rapid}, the latter can be 
interpreted as a double form factor. This quantity,  usually
called effective form factor (eff), can be   defined 
as 
follows:

\begin{align}
\label{eff1}
 F_2(k_\perp^2) = \int_0^1 dx_1~\int_0^{1-x_1} dx_2~F_{q_1 
q_2}(x_1,x_2,{\bf k_\perp} )~.
\end{align}
 For unpolarized dPDFs, the eff does not depend 
on the direction of ${\bf k_\perp}$.  The above definition is general 
and  also valid for many-body systems. Moreover, one should notice 
that 
the normalisation of the LF wave function relies in the condition:  
$F_2(0)=1$. Physically, the latter ensures that the Fourier Transform 
(FT) of the eff can be interpreted as the 
probability 
of finding two partons with a given transverse distance ${\bf d_\perp}$.
This quantity is indeed the conjugate variable to ${\bf k_\perp}$. Let 
us 
stress that  
 a pre-factor in Eq. 
(\ref{eff1}), depending on the kind of hadron, could appear according 
to the 
dPDF sum rules \cite{gaunt1}.
In the meson case, where only a $q \bar q$ state is considered, the 
eff reads:

\begin{align}
 F_2(k_\perp^2)&= \int_0^1 dx ~f_2(x, {\bf k_\perp})
= {1 \over 2} \sum_{h,\bar h} 
 \int_0^1 dx~   \int {d^2{\bf k_{1\perp}} 
\over 2 (2 
\pi)^3}~  \psi_{h, 
\bar h} (x,{\bf k_{1 \perp}}  ) \psi_{h, 
\bar h}^{*} (x,{\bf k_{1 \perp}+k_\perp}  )~.
\label{eff2}
\end{align}
The above quantity will be calculated in the next sections and compared 
with that extracted from the lattice QCD~ \cite{lattice}.

\subsection{
\label{IIA}An approximation in terms of one body quantities}

In order to phenomenologically estimate the magnitude of the DPS cross 
section in proton-proton collisions,  an 
approximate relation between GPDs and dPDFs is often assumed in 
experimental analyses~
\cite{blok1,blok2}.
In fact, 
by introducing
 a complete set of states 
in  the correlator (\ref{f2c1}) 
and keeping only the mesonic contribution, one  gets

\begin{align}
\nonumber
  F_{q_1 \bar q_2}(x_1,x_2, {\bf k_\perp} )&\sim
 {P^+ \over 4 } \int d^2 {\bf y_\perp} 
e^{-i 
{\bf 
y_\perp} \cdot {\bf k_\perp} }\int dy^- 
 \int dz^-_1dz_2^- \int \dfrac{dP'^+d^2 {\bf P'_\perp}  
}{2(2\pi)^3 P'^+} {e^{-i x_1P^+ z^-_1-i 
x_2P^+ z^-_2} \over (2\pi)^2 }
\\
& \times
  \langle M, {\bf 0}| \mathcal{O}_{q_1}(0,z_1) |M,{\bf 
P'_\perp}\rangle \langle M, {\bf P'_\perp}|
\mathcal{O}_{\bar q_2}(y,z_2)|M, {\bf 
 {\bf 0}} \rangle \Big|_{y^+ =z^+_1=z_2^+=0}^{{\bf z_{1\perp}= 
z_{2\perp}  }=0  
}~.
\label{f2c1a}
\end{align}
By using the strategy already 
discussed in the previous section, one finally finds:

\begin{align}
 \label{f2ma1}
 F_{q_1 \bar q_2}(x_1,x_2,{\bf k_\perp}  ) &\sim H_{q_1}(x_1,{\bf 
k_\perp} 
)H_{\bar q_2}(1-x_2,-{\bf 
k_\perp} 
)~,
\end{align}
where $H_q(x, {\bf k_\perp})=H_q(x, \xi=0,{\bf k_{\perp}})$, is the 
meson 
GPD at 
zero 
skewness (see Refs. \cite{dihel_gpd,gpd} for  useful reports on GPDs).
Let us mention that the above expression has been tested, in the proton 
case, by using a LF quark model~ \cite{noij2}.
{  The integral over $x_2$ of Eqs. (\ref{fdiv}) and (\ref{f2ma1}) leads 
 to }

\begin{align}
 \label{f2ma}
 f^M_{2}(x,{\bf k_\perp}  ) \sim  f^M_{2,A}(x,{\bf k_\perp})=
H^M(x,{\bf k_\perp} )F^M({\bf k_\perp} 
)~,
\end{align}
where $F^M( {\bf k_\perp} )$ { is} the standard  e.m. form 
factor of the meson M. We denote   the meson dPDF, 
evaluated within 
the above ansatz, as $f^M_{2,A}$.
 The difference between the full calculation of the dPDF and its 
approximation can be interpreted as the sign of the presence of 
correlations not encoded in 
one-body quantities, such as GPDs and ffs. A dedicated numerical 
section
about 
the impact of correlations in dPDFs   will 
follow.
Let us mention that an approximated expression of the first moments of 
the dPDF 
can be also obtained. In this case, the integration over $x$ of the 
expression 
(\ref{f2ma}) leads to:

\begin{align}
\label{appmom}
 F_2(k_\perp^2) &\sim \int_0^1dx~ H(x,{\bf k_\perp})F(k_\perp^2)=
F({ k_\perp^2})^2~.
\end{align}
A similar ansatz has been tested
in the 
lattice investigation of Ref. \cite{lattice}. In Eq. (\ref{appmom}),
  the relation between the GPDs and ffs has been used  
\cite{dihel_gpd}:

\begin{align}
\label{appff}
 F(k_\perp^2)= \int_0^1dx~H(x,{\bf k_\perp})~
\end{align}
The above form factor can be described in terms of 
the LF wave function. For a meson described by the first Fock state, 
one gets 
the following expression \cite{Br1,drell,west}:

\begin{align}
\label{ffc}
 F(k_\perp^2) =  \int_0^1 dx~ \int {d^2 {\bf k_\perp} \over 16 
\pi^3 } 
\psi^* 
(x,{\bf 
k_{ \perp}}  ) \psi (x,{\bf 
k_{ \perp}} +(1-x ){\bf k_\perp} )~, 
\end{align}
where $k_\perp = |{\bf k_\perp}|$. The approximation Eq. 
(\ref{appmom}) will be numerically tested by means of holographic quark 
models.

\subsection{\label{IIB}   The effective cross section }
In this  section, the so 
called 
effective 
cross section, $\sigma_{eff}$ \cite{Calucci:1999yz},
a relevant 
 observable { for { DPS}  studies}, is introduced. 
This quantity 
 is defined as the ratio of the product of two single parton 
scattering process 
cross sections to the DPS one  with { the } same final states. Usually
$\se$
{  is extracted from data by using model assumptions, such as the 
factorisation of dPDFs in terms of PDFs.}
Experimental analyses, for proton-proton collisions, have been already 
compared with 
quark model calculations of $\se$~ 
\cite{noij3,Traini:2016jru,noiplb,noiW,Gaunt:2010pi}.
Let us mention that in Ref. \cite{Traini:2016jru} an  AdS/QCD soft-wall
model for the proton has been used to calculate this quantity.
The common
feature, pointed out
in  Refs. \cite{noij3,Traini:2016jru,noiplb,noiW}, is the 
dependence of $\se$ 
 on the longitudinal momentum 
fractions carried by { the}  acting partons.
This behaviour is interpreted as the effects of
 non 
trivial 
double parton correlations. 
Although no experimental analyses for the extraction of $\se$ for 
meson-meson collisions are available, in the present investigation the 
above quantity 
will be evaluated to make predictions for DPS processes involving 
mesons. Let us 
mention that for the pion 
case, the estimate of $\se$, shown in Ref. \cite{noipion}, has been 
used 
in the experimental investigation of Ref. \cite{Koshkarev:2019crs}.
The general definition of this quantity is
\cite{bansal}:

\begin{eqnarray}
\label{pocket}
\sigma_{eff} = { m \over 2 }
{ \sigma_A^{pp'} \sigma_B^{pp'} \over 
\sigma^{pp}_{double}
}~.
\end{eqnarray}
$m$ is a process-dependent combinatorial factor: $m= 1$
if $A$ and $B$ are identical and $m=2$ if they are different. 
$\sigma_{A(B)}^{pp'}$ is the differential
cross section for the inclusive process $pp' \rightarrow A (B) + X$.
 As a first approximation for experimental analyses, 
$\se$ is considered 
rather independent from the flavors of the partons, the final states of 
 the 
processes and the experimental kinematic conditions.
However,
 recent studies on quarkonia production 
suggest that this ansatz might be violated~ 
\cite{data12}. 
Due to the lack of experimental data for meson-meson DPS processes, in 
the 
present study we calculate the  mean value of $\se$ 
in order to discuss its geometrical interpretation \cite{rapid}:

\begin{align}
 \label{sieffa}
 \overline{ \sigma}_{eff}  = 
 {1 \over  { \displaystyle \int} { d^2 {\bf 
k_\perp} \over  
(2\pi)^2 
} 
 F_{2}^\pi( {\bf k_\perp} )F_{2}^\pi(- {\bf k_\perp} )  }~.
\end{align}
Let us mention that if double parton correlations could be neglected,
then $\se = \mse$. Anyhow,
the above expression  encodes unknown
non perturbative insight on 
the hadronic structure, such as the geometrical 
information on the system.

\subsubsection{On the geometric interpretation of $\overline{ 
\sigma}_{eff}$}

As already pointed out in the previous section,
due to the lack of experimental information on double 
parton scattering 
processes, in  particular for meson targets, 
calculations on $\overline{ \sigma}_{eff}$ could be relevant to make 
predictions, such as the one of Ref. 
\cite{Koshkarev:2019crs}. In this 
scenario, 
the interpretation of
$\sigma_{eff}$, in terms of geometrical properties of the incoming 
hadron, is fundamental.
 To this aim, in this section, 
 we explore an intuitive relation between 
$\overline{ \sigma}_{eff}$ and the
mean partonic distance between two partons acting in a DPS process. 
This study has been discussed in detail in Refs. \cite{noij3,rapid}.
The procedure is somehow 
similar to that applied in the 
case of elastic processes, where the e.m. form factor, extracted 
from the 
relative cross section,
can be related to
  the charge/magnetic radius. However, since $\mse$ 
depends on
the integral over $k_\perp$ of the product of two effs  
(\ref{sieffa}),  a direct extraction of the eff is precluded. 
Nevertheless, basic probabilistic properties of the FT of this 
quantity allow 
to relate $\mse$ to the main partonic transverse 
distance 
between 
two partons $\langle d^2_\perp \rangle$. 
The effective form factor \cite{noiplb}, for a generic system, can be 
indeed
defined as follows:

\begin{align}
 \label{aff2}
 F_{2}(k_\perp) = \int d^2{\bf d_\perp} ~ e^{i {\bf k_\perp}\cdot  
{\bf 
d_\perp}     } \rho(d_\perp)~,
\end{align}
 being $\rho(d_\perp)$ the two-body density of the system for two 
particles 
whose distance in the transverse plane is 
${\bf d_\perp}$.
 Thanks to this relation, one  finds:

\begin{align}
 \label{2dist}
  \langle d^2_\perp \rangle 
\simeq 
-4 { d 
F_2(k_\perp) \over 
dk_\perp^2} \Bigg|_{k_\perp=0}~.
\end{align}
{ The above expression, first introduced in 
Ref. \cite{rapid} and  applied in the Lattice QCD analysis of Ref. 
\cite{lattice}, is a generalisation of the standard relation between 
the mean square radius of the the proton and its relative form form 
factor. Let us remark that we are considering unpolarized quarks 
in an
unpolarized hadron; thus the eff  depends on  $| \vec k_\perp |$. 
Details on this relation can be found in Ref. 
\cite{rapid}}.
Due to this connection between the effective form factor and the mean 
distance 
of two partons,  
one can  relate $\mse$  (\ref{sieffa}) 
to the above 
quantity.  Here and in the following we refer to $\mse$ as the 
geometrical effective cross section. The latter is indeed a process 
independent constant depending only on the functional behaviour of the 
eff.
In Ref. \cite{rapid}, the relation between the numerical 
value of  $\mse$ and the partonic distance has 
been properly understood. Here   the main outcome of Ref. \cite{rapid}
is shown.
By considering the definition of $\bar \sigma_{eff}$ 
(\ref{sieffa}) and 
the probabilistic interpretation of the FT of the eff, one can show 
that the main partonic distance (\ref{2dist})
lies in a range depending on $\bar \sigma_{eff}$  as follows:

\begin{align}
 \label{ine}
\dfrac{\bar \sigma_{eff}}{3 \pi} \leq \langle d^2_\perp \rangle \leq 
\dfrac{\bar \sigma_{eff} }{\pi}
\end{align}

Such a result is extremely useful to get some information on the 
geometrical structure of an hadron once some data on $\sigma_{eff}$ are 
collected. Since in the present analysis
the mean partonic distance will be calculated within quark models and 
compared to that obtained from the lattice QCD, the above inequality 
(RC) will be  tested. Let us remind that in the proton 
case the RC inequality  has been verified by using all quark models and 
ansatz of dPDFs 
at our disposal~ \cite{noij3,rapid}. Furthermore, in the pion 
case, the above relation has been also validated by the NJL model~ 
\cite{njl2}.

\section{Lattice analysis of moments of dPDFs}
\label{III}
In this section, we briefly recall the main formalism introduced in 
Ref. \cite{lattice}. Here,
the expectation for the two-current distribution, a quantity related 
the first moment of the pion dPDF, has 
been evaluated within the lattice framework. In momentum space, this 
quantity reads:

\begin{align}
 \label{lattice1}
M(q^2)= \int d^3y~
e^{i \vec y \cdot \vec q} 
\langle \pi, p| {\cal O}(y) {\cal O}(0) | \pi, p \rangle 
\Big |_{y_0=0}~.
\end{align}
The main differences with respect to the light-cone derivation of the
dPDF are: $i)$ the gamma matrix considered  in Eq. 
(\ref{lattice1})
 is 
$\gamma^0$, instead of $\gamma^+$ in Eq. (\ref{op}); $ii)$ 
the distance between the quark field operators $y$ is chosen along 
the condition
$y_0=0$, instead of $y^+=0$, see Eq. (\ref{f2c1}). 
 However, as discussed in the Appendix A, kinematic 
corrections, due to the choices
of the gamma matrix and
the separation condition,  
can be neglected in the infinite momentum frame (IMF), i.e. the natural 
reference system where a partonic description of hadrons can be 
provided. Therefore numerical comparisons, between lattice and quark 
models 
calculations, are allowed in this frame. However,
one of the main consequences of  the conditions $i$ and $ii$ 
is the  frame dependence of 
 numerical evaluations of Eq. (\ref{lattice1}) within the 
lattice approach \cite{lattice}. In this framework, moments of dPDFs 
depend upon the 
 the pion momentum $\vec p$. This feature will be 
explicitly relevant in the analysis of the approximation  
(\ref{appmom}).
In the next section,
 the comparisons of 
the double parton correlations effects, highlighted in the analysis of 
Ref. 
\cite{lattice}, with those addressed in constituent quark model 
calculations, will be presented. To this aim the
 lattice data, we are interested for, are here shown. For simplicity, 
all
distributions will be evaluated in 
momentum space.

\subsection{The pion form factor}
The standard electromagnetic (e.m.) ff, necessary 
to test the approximation  (\ref{appmom}),
has been directly fitted  from 
lattice results. The expression reads:

\begin{align}
 \label{latticeff}
F_L(Q^2) = \dfrac{1}{\left[1+\dfrac{Q^2}{M^2}   \right]^n }~,
\end{align}
where the parameters leading to a good fit with lattice data are:
$M=0.872 \pm 0.016$ GeV and $n=1.173 \pm 0.069$ (configuration $A$) or 
$M=0.777 \pm 0.012$ GeV 
and $n=1$ (configuration $B$) \cite{lattice}.
 As one can observe in the left panel of Fig. \ref{lattice12}, 
differences between the two configurations are minimal. 
 In 
the present investigation, the A
configuration has been used as benchmark for further comparisons.

\subsection{Effective form factor}
As already pointed out, the first moment of a dPDF is the eff 
(\ref{eff1}).
 However, within the lattice framework,  one finds that 
$M(0)=-2 m_\pi$, thus, following the procedure of Ref. \cite{lattice}, 
the eff is properly defined as follows

\begin{align}
 \label{latticeff2}
 F_{2L}(q) \equiv \dfrac{M(q)}{-2 m_\pi} =
 \dfrac{1}{\left[ 1+\langle 
d^2 \rangle \dfrac{q^2}{6n}  \right]^n}~,
\end{align}
here the parameter $n$ is the same of that of $F_L$ of Eq. 
(\ref{latticeff}). 
Let us stress that the FT of the above expression has  the 
probabilistic interpretation shown in Eq. (\ref{aff2}). In fact,
within the above 
functional form,   the 3-dimensional mean distance between 
the 
two partons is: $\sqrt{\langle d^2 
\rangle} = 1.046$ fm \cite{lattice}. Let us remark that since fort the 
moment being only unpolarised quarks in the unpolarised pion, then 
w.r.t. the definition Eq. (\ref{2dist}), $\langle d^2 \rangle = 3/2 
\langle d^2_\perp \rangle$.
As deeply discussed in Ref. \cite{lattice}, the quantity $M(q^2)$, 
 has been numerically evaluated in the pion rest 
frame, i.e. $\vec p=0$. However, as previously mentioned, a comparison 
between lattice data and quark model calculations of dPDFs is possible 
in the IMF 
(see discussion in Ref. \cite{lattice}). 
Thus in order to proceed with the present study, it is necessary to 
realise that the IMF can be approximately mimicked in the kinematic 
regions where $q^2 << 
m_\pi^2$.  We recall that in the 
lattice QCD analysis  \cite{lattice} the pion mass is fixed to 
be 
$m_\pi=0.3$ GeV.

\subsection{ An approximation for the moment of dPDFs in lattice QCD}
\begin{figure*}
\includegraphics[scale=0.80]{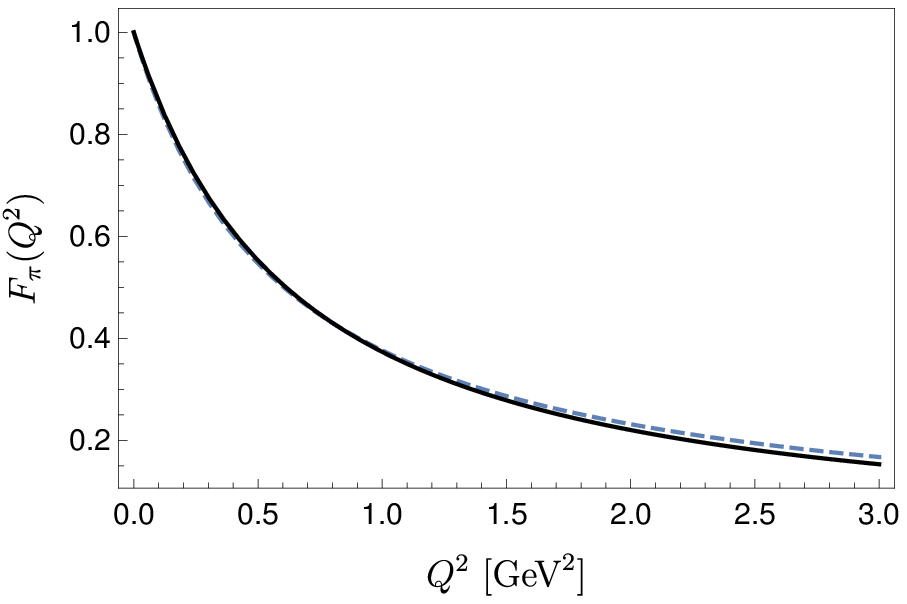}
\hskip 0.5cm \includegraphics[scale=0.80]{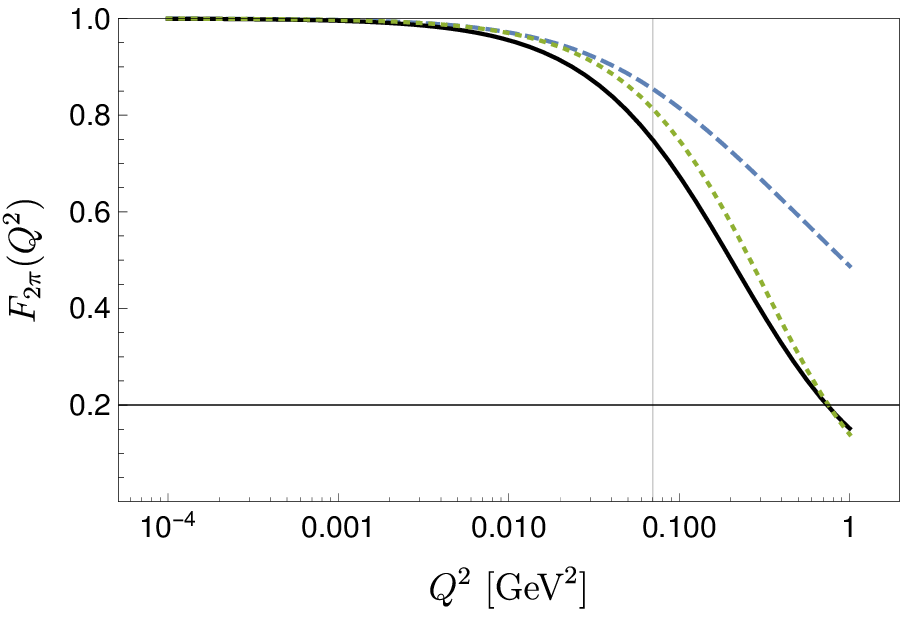}
\caption{\footnotesize  \textsl{Left panel: the pion e.m. form factor 
evaluated 
within the lattice framework, see Eq. (\ref{latticeff}). Full black 
line for 
the configuration $A$ and dashed blue line for the configuration $B$. 
Right panel: 
comparison between the lattice eff (\ref{latticeff2}) (full black line) 
and  
its 
approximation  (\ref{latticef2a})  (dashed blue line).
Dotted green lines stand for the approximation  (\ref{latticef2a})
evaluated in the IMF, see  Eq. (\ref{latticef2aIMF}). }}
\label{lattice12}
\end{figure*}

As mentioned in Sect. \ref{IIA},  a direct measure of the impact of 
unknown DPCs is the discrepancy between the eff 
and 
its approximation in terms of the e.m. form factor, see Eq. 
(\ref{appmom}).
To this aim, the procedure
discussed in  Sect. \ref{IIA}
has been considered also in the lattice 
analysis~ \cite{lattice}. However,  since in this framework frame 
dependent 
effects appear, the following result is 
obtained:

\begin{align}
 \label{latticef2a}
F_{2L}(Q^2) \sim  \dfrac{(m_\pi+E_q)^2}{4 m_\pi E_q} \Big[ F_L(2m_\pi 
E_q-2m_\pi^2) \Big]^2 =\bar F_L(q)^2~.
\end{align}
As already  explained \cite{lattice}, the above result comes 
from 
the procedure discussed in Sect. \ref{IIA} but using the lattice 
conditions described in the first part of Sect. \ref{III}. In 
particular, 
 the above expression has been obtained in the 
pion rest frame, i.e. $\vec p=0$ \cite{lattice}. As one can see, the 
approximation  (\ref{latticef2a})
 is different from 
that derived within the light-cone formalism, see  Eq. 
(\ref{appmom}). However, it is remarkable that in the IMF the standard 
expression Eq. (\ref{appmom}) is 
recovered from Eq. (\ref{latticef2a}). In fact, by
 replacing the 
 pion energy at rest  with that of a moving target with  an extremely 
large  momentum 
$\vec p$:
 $m_\pi \rightarrow E_p= \sqrt{m_\pi^2+p^2}$,  
one gets:

\begin{align}
 \label{latticef2aIMF}
F_{2L}(Q^2) \sim \bar F_L(Q^2)^2=   F_L(Q^2)^2~,
\end{align}
for $q^2 << p^2$.
This is exactly the result found by following the standard strategy 
discussed in Sect. \ref{IIA}, see Eq. (\ref{appmom}). Let us remark 
that such a conclusion can be also reached
 by imposing $q^2 
<< 
m_\pi^2$ (see discussion on Ref. \cite{lattice}).
A direct consequence of this approximation is the 
relation between the mean partonic distance and the mean pion radius.
In fact by using that:

\begin{align}
\label{ratio1}
 \langle r^2 \rangle = -6 \dfrac{d F_L(Q)}{d Q^2} \Big|_{Q^2=0} = -6 
\dfrac{d \bar F_L(Q)}{d Q^2} \Big|_{Q^2=0}~,
\end{align}
and by considering the relations Eqs. 
(\ref{latticef2a}-\ref{latticef2aIMF}) between the eff and the e.m. 
ff, one finds that
 $\langle d^2 \rangle \sim 2 \langle r^2 
\rangle$. Numerical values, obtained within the lattice techniques 
\cite{lattice}, for $\langle d^2 \rangle $ and $\langle r^2 
\rangle$, immediately show that $ \sqrt{ \langle d^2 
\rangle}=1.046 \ne \sqrt{2\langle r^2 \rangle}=0.85$ fm. As one can 
observe, correlations effects prevent a simple relation between these 
two quantities.
Let us stress that since both $\langle d^2 \rangle$ and $\langle r^2 
\rangle$ depends on the small $Q^2$ behaviour of effs and ffs,
they are rather 
independent on the chosen frame. Such a feature has been confirmed by 
numerical calculations of each sides of Eq. (\ref{ratio1}), see Ref. 
\cite{lattice}.
In addition, details on the impact of DPCs can be obtained 
by comparing both sides of
Eqs. (\ref{latticef2a}-\ref{latticef2aIMF}) 
as a function of $Q^2$.
As one can see in the right panel of Fig. \ref{lattice12}, 
the presence 
of 
correlations prevents  a simple description of the eff in terms of 
standard ff. See the difference between the full line (left hand side 
of Eq. (\ref{latticef2a})) and the dashed line (right hand side of Eq. 
(\ref{latticef2a})).
However, in the lattice framework, DPCs mix with frame dependent 
effects, thus,  in the right panel of Fig. \ref{lattice12} we also plot 
 the right hand side of  Eq. 
(\ref{latticef2aIMF})
i.e. the  approximation 
in the infinite momentum 
frames (dotted lines). The comparison between dotted and dashed lines, 
provides a  numerical estimate of the region where frame dependent 
effects are minimal.
One can observe that 
calculations 
 obtained in the pion rest 
frame are close to those obtained in the IMF up to $q^2 < m_\pi^2 $, 
as expected. From this check one can deduce  that a 
comparison, between 
lattice data  and predictions of holographic QCD models, are 
allowed for $q^2 < 0.07$ GeV$^2$.
Before closing this section, the explicit expression of the pion eff
(\ref{latticeff2}) has been used to evaluate the geometrical effective 
cross section: $\bar \sigma_{eff}=26.3$ mb.   Since this result 
correspond to the case of pions in 
their rest frame,
this numerical result is rather useless for experimental analyses. On 
the contrary, $\langle d^2 \rangle$ is almost frame independent. In 
fact, this quantity depends on the behaviour of the eff at $k_\perp 
\sim 0$ (see discussion in Ref. \cite{lattice}). Therefore,
 by inverting 
the RC inequality (\ref{ine}), one can estimate a range of possible 
$\mse$ once  the value of $\langle d^2 \rangle$ is established:

\begin{align}
 \label{ine2}
  \dfrac{2 \pi }{3}  \langle d^2 \rangle \leq \mse \leq  \langle 
d^2 \rangle 3 \pi~.
\end{align}
 From  the above expression,
an allowed 
 range of $\mse$, valid also in the IMF, can be estimated.
Starting from the lattice data  $\langle 
d^2 \rangle=1.046$ fm$^2$, one gets:

\begin{align}
\label{uff}
 22.9 ~ [\mbox{mb}] \leq \mse \leq 68.7~  [\mbox{mb}]~.
\end{align}
{ The relevance of the above result relies on its frame 
independence. Indeed, while the eff, extracted by Lattice collaboration 
depends on 
a given frame, the value of $\langle d^2 \rangle$ does not. In fact, 
as one might notice in Eq. (\ref{2dist}), this quantity depends on the 
small $k_\perp^2$ behaviour of the effective form factor. Therefore, 
$\langle 
d^2 \rangle$ is related to kinematic regions where frame dependent 
effects are relatively small. Thanks to this feature, one might 
conclude that the inequality
(\ref{uff}) is frame independent too. In this scenario, even if
a direct experimental prediction  from lattice data 
cannot be 
safely obtained, thanks to the above procedure an hint on the amount of 
  $\sigma_{eff}$ can be provided. Let us stress that for the moment 
being such a quantity is related to  an hypothetical DPS process 
involving pions. }
From this general results of the lattice QCD,  one can conclude that 
the 
mean 
value of $\se$ for a pion-pion collision is 
bigger then that extracted in  proton-proton collisions. {
Since, as shown in Eq. (\ref{pocket}), $\sigma_{eff}$ estimates the 
ratio between the DPS process to the product of two SPS processes, 
the result Eq. (\ref{uff}) can be physically interpreted 
 as a suppression of the DPS 
contribution, w.r.t. the SPS one, bigger in pion then in the proton.
This outcome could guide future phenomenological analyses of DPS off 
mesons. }

\section{The pion dPDF within the holographic QCD}
\label{IV}
In this section, details on the constituent quark models adopted to 
investigate basic feature of pion dPDF will be presented.
In particular, we are interested in the mesonic wave function 
calculated 
by 
using  different Light-Front holographic QCD  models. 
The first w.f. described in this section has been introduced in Ref. 
\cite{Br1}.
The pion dPDF has been evaluated for the first time within this model 
in Ref. 
\cite{noipion}. However, since the 
aim of this analysis is to provide a first comparison with lattice 
data, the  pion wave function has been also evaluated by improving 
that of 
Ref. \cite{Br1}. To this aim,  we also considered 
the model where dynamical spin effects have been taken into account~
\cite{spin}. In addition, in order to include other fundamental 
phenomenological 
effects, such as the Regge trajectory of the $x$-dependence of PDFs, 
the 
model of Ref. \cite{universal} has been also adopted.

\subsection{Pion in AdS/QCD I: The original version}
In this section, we discuss the calculation of the pion dPDF 
evaluated within the 
model  described in
 Refs. \cite{Br1,Br2}.
Since the w.f. obtained in this scenario can be considered as the 
starting point for any 
further 
implementations, here and in the following, we refer to it as the 
``original'' model. Indeed, it 
can reproduce basic properties of the meson spectroscopy and structure 
functions.
In  momentum space representation, the pion wave function
 reads \cite{Br1}:

 \begin{align}
\psi_{\pi o} (x,{\bf k_{1 \perp}}  )\propto {4\pi 
\over  \kappa_o \sqrt{x(1-x)} } 
e^{- { {\bf k^2_{1\perp} } \over x(1-x) 2\kappa^2_o  }  }~,
\label{ori}
\end{align}
 
being $\kappa_o = 0.548$ GeV fixed to reproduce the Regge 
behaviour 
of the mass spectrum of mesons. Moreover, in
 order to 
include a dependence on the quark masses, the wave function has
 been  written 
in 
terms of the invariant mass~\cite{Br2}:

\begin{align}
 M^2 = \sum_i {m_i^2 + {\bf k_{i \perp} }^2 \over x_i } = {m^2 + {\bf 
k_{1\perp} }^2  \over x(1-x)}~,
\end{align}

where 
 $m=m_1 \sim m_2$, $x=x_1$, $x_2 = 1-x_1$ and 
${\bf 
k_{2\perp} } = - {\bf k_{1\perp}  } $. In this scenario the pion 
wave function now 
reads:

  \begin{align}
\label{pionwfori}
\psi_{\pi o} (x,{\bf k_{1 \perp}}  )= A_o {4\pi 
\over  \kappa_o \sqrt{x(1-x)} } 
e^{- { {\bf k^2_{1\perp} } + m_o^2 \over x(1-x) 2\kappa^2_o  }  }~.
\end{align}
 
 The mass parameter is usually chosen to be $m_o \sim 0.33$ GeV~
\cite{15sp}. The  constant $A_o$ is  fixed by 
the following 
normalisation condition:

 \begin{align}
 \label{bo}
 \int_0^1 dx~\int {d^2 {\bf k_{1\perp} } \over 16 \pi^3  } 
 |\psi_{\pi o} 
(x,{\bf 
k_{ 1\perp}}  )  |^2 = 1~.
\end{align}
 Within this approach the
 dPDF expression 
for the pion can be analytically found \cite{noipion}:

\begin{align}
 \label{dpdfpiono}
 &f_2^{\pi O}(x, {\bf k_\perp} ) = A_o^2 
e^{- { 4m_o^2+  k_\perp^2 \over 4 
\kappa_o^2 
x (1-x) }  }~.
\end{align}

\subsection{Pion in AdS/QCD II: Dynamical spin effects in 
holographic QCD}
In Ref. \cite{spin}, dynamical spin effects have
 been included 
into the holographic 
pion wave function  in order to predict the mean charge 
radius of the pion and its ff without including high Fock states in 
the meson expansion  (\ref{state}). To account these contributions, 
let us 
promote the function appearing in Eq. (\ref{pionwfori}) as an 
helicity dependent
quantity, i.e.

\begin{align}
\label{wfspin}
 \psi_{\pi s}(x, {\bf k_\perp} ) = S_{h \bar h}(x, {\bf 
k_\perp}) \psi_{\pi o}(x, {\bf k_\perp} )~.
\end{align}
Without going into details, let us discuss only
 the main 
outcomes of Ref. \cite{spin}. The spin  operator reads:

\begin{align}
 S_{h \bar h}(x,{\bf k_\perp} ) = \left[ A m_\pi^2 +B 
\left(  { m_o m_\pi 
\over x(1-x)  }     \right)        \right] (2 h) \delta_{-h \bar h}
+ B \left[  { m_\pi k_\perp e^{ i (2h)\theta_{k_\perp} } \over x(1-x) }
\right] \delta_{h,\bar h}~,
\end{align}
where ${\bf k_\perp}=k_\perp e^{i \theta_{k_\perp}}$.
The original model, described in the previous section is restored 
for $B=0$ and 
$A = 1/m_\pi^2$, i.e.:

\begin{align}
 \left\{      \begin{array}{l}
S_{h\bar h}(x, {\bf k_\perp}) 
\underset{B=0}{\overset{A=1/m_\pi^2}{\longrightarrow}}  {1 \over 
\sqrt{2} } (2h) \delta_{-h 
\bar h}\\
\\
\underset{h,\bar h}{ \sum} |S_{h \bar h}|^2 =1
              \end{array}
    \right.
\end{align}

Where the last condition  ensures the normalisation of the pion wave 
function. 
Let us call the dPDF evaluated within the present 
model,  $f_2^{\pi 
AB}$, where $A$ and $B$ 
can assume different values.
By following Ref. \cite{spin}, we consider two configurations, i.e. 
$A=B=1$ and $A=0,~ B=1$.
In particular,
the w.f. entering   Eq. (\ref{wfspin})
is the same of that obtained within the original model discussed in the 
above 
section, see Eq. (\ref{pionwfori}). However, in order to recover 
phenomenological predictions, the parameter entering the w.f. Eq. 
(\ref{pionwfori}) is $\kappa_0=0.523$ GeV~\cite{spin}.
Also in this case,  analytic expressions for dPDFs can be found:

\begin{align}
\label{dpdf_pi11}
 f_2^{\pi11}(x, {\bf k_\perp}) &= \frac{0.0415248}{x^2 (1 - x)^4}  
e^{\frac{0.91398~ {k^2_\perp}+0.39813}{x ( x-1)}}
  \left. \Big[-12.7551 ~{k^2_\perp} (1 - x)^2+ x^6-4 x^5 \right.
 \\
 \nonumber
& \left.  -12.6698
   x^4+52.0095 x^3-49.4534 x^2 \right.
  \left.+7.5576 x+5.55612\right. \Big]~,
\end{align}

for the case where $A=B=1$ and

\begin{align}
\label{dpdf_pi01}
 f_2^{\pi01}(x, {\bf k_\perp}) &= -\frac{0.64739}{(x-1) x^2 ( x-1)^3} 
e^{\frac{0.91398~ {k^2_\perp}+0.39813}{x (x-1)}}
 \left. \Big[0.91398~ {k^2_\perp} (1 - x)^2+x^4-3 x^3 \right.
\\
\nonumber
& \left. + 2.60187
   x^2-0.203741 x-0.39813\right. \Big]~,
\end{align}

for the case where $A=0$ and $B=1$. Here and in the following we refer 
to this model as the ``dynamical spin model''.

\subsection{Pion in AdS/QCD III:   A universal wave function}
In this last part of this section, a new and promising pion wave 
function, obtained from the holographic correspondence, will be 
presented~ \cite{universal}. In this case, the basic idea is 
to 
consider the most general analytic structure of GPDs, obtained within 
holographic QCD, and then incorporate the Regge trajectories for small 
$x$ in PDFs. In this  procedure, the mathematical structure preserves
 the poles of the ff in the physical region. Here and in the 
following we indicate this model as the ``Universal model'' (UM). Let 
us 
here just 
remind the main outcomes  of Ref. \cite{universal}. 
Within this model, the
effects of two Fock states in the hadron  expansion (\ref{state}) 
are 
considered: the valence configuration $|q \bar q \rangle$ and the 
$|q\bar 
q 
q \bar q \rangle$ contribution. These two different states are 
addressed with the 
index 
$\tau=2$ and $\tau=4$, respectively. A remarkable result shown in ~ 
\cite{universal} is that  nucleon and pion PDFs, GPDs and ffs can be 
described 
within the same model. Of course, free parameters are chosen to 
describe ffs, PDFs and hadron spectroscopy at the same time. The w.f., 
related 
to a given $\tau$ state can be effectively expressed as follows:

\begin{align}
 \label{effwf}
\psi^\tau_{eff}(x, {\bf k_\perp}) = 8\pi \dfrac{\sqrt{q_\tau(x)f(x) }  
}{1-x}Exp{ \left[ \dfrac{2 f(x) }{(1-x)^2} {\bf k_\perp^2}  \right] } 
~,
\end{align}
where here $q_\tau(x)$ is the $\tau$ contribution to the pion PDF. The 
analytical structure of this quantity is fixed by the 
holographic QCD approach:

\begin{align}
 q_\tau(x) = \dfrac{1}{N_\tau} \big[   1-w(x)^{\tau-2}\big]
w(x)^{-\frac{1}{2}} 
w'(x)~,
\label{unt1}
\end{align}
where:

\begin{align}
\label{unt2}
 w(x)&=x^{1-x}e^{-a(1-x)^2}
\\
\label{unt3}
f(x)&=\dfrac{1}{4 \lambda}\left[(1-x)\log\left(\dfrac{1}{x}  \right)+a 
(1-x)^2   \right]
\\
\label{unt4}
N_\tau &=\sqrt{\pi} \Gamma(\tau-1)/\Gamma(\tau-1/2)~.
\end{align}
Thanks to this choice the 
Regge trajectory is correctly reproduced. Moreover,
the parameters 
$a$ and $\lambda$ have been phenomenologically fixed by fitting the 
mesonic mass spectrum and the e.m. form factor. Results are 
found for  $a=0.531$ and $\kappa=\sqrt{\lambda} 
= 0.548$ GeV. 
In Ref. \cite{universal}, the authors fixed the weight of the two Fock 
states, $\gamma$, 
contributions by using  the pion moment of PDFs:

\begin{align}
\label{PDFtau}
 f_1^{\pi U}(x) = (1-\gamma) q_{\tau=2}(x)+\gamma q_{\tau=4}(x)~.
\end{align}
In particular $\gamma = 0.125$ \cite{universal}. 
Let us point out that the wave function of the $\tau=4$ state is 
computed only to calculate PDFs. Thus, the dependence of the latter 
upon 
the other two particle momenta is integrated out. 
Thanks to all these ingredients, the pion 
dPDF,
$f_2^{\pi U}(x,k_\perp)$, can be 
evaluated. In the next section numerical results will be discussed.
{ Let us mention that this model represents an important 
improvements w.r.t. the original one. Indeed, in this scenario the 
Regge 
behaviour at small $x$ has been properly  included together with pole 
structure of the form factor. In addition, let us remark that a 
contribution of higher Fock states to the hadron PDF has been 
effectively incorporated. As it will be discussed later on, such a 
feature is quite relevant in the present analysis.}

\section{Numerical Results}
\label{V}

In this section, numerical results
of the calculations of dPDFs, within  holographic models,
 will be presented. 
In 
particular, we will mainly focus on 
quantities which allows to qualitatively estimate the impact of non 
perturbative double parton correlations, not directly accessible via 
one-body distributions.

\subsection{Calculation of dPDFs}
\begin{figure*}
\includegraphics[scale=0.80]{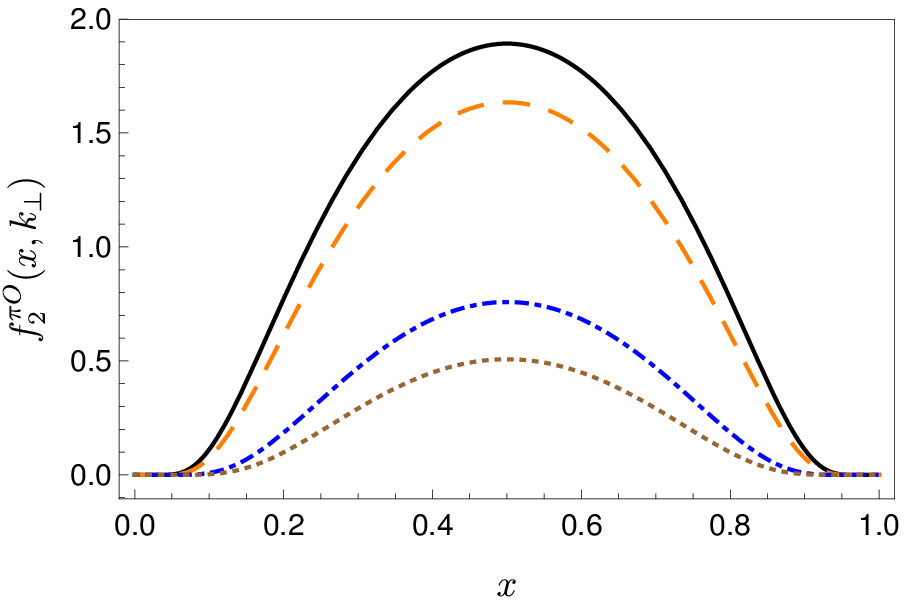}
\hskip 1cm \includegraphics[scale=0.80]{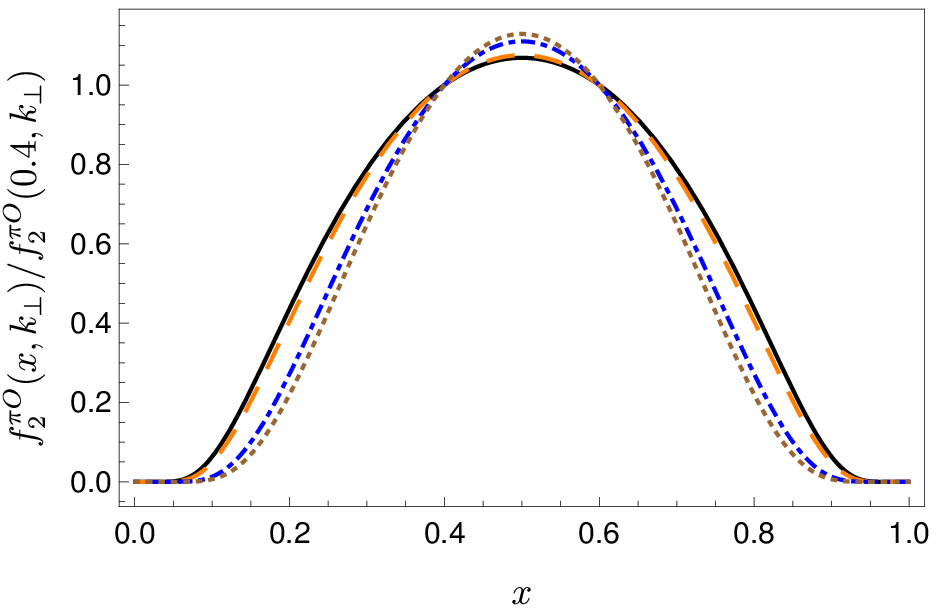}
\caption{\footnotesize  \textsl{Left panel: dPDFs evaluated
within the original model 
\cite{Br1} and 
addressed for 
different values of $k_\perp$. Full black line $k_\perp=0$ GeV, dashed 
orange line 
$k_\perp=0.2$ GeV, dot-dashed blue line $k_\perp=0.5$ GeV and dotted  
brown line 
$k_\perp 
= 0.6$ GeV. Right panel: same of the left panel for the ratio Eq. 
(\ref{rk}). }}
\label{dpdfpo}
\end{figure*}

\begin{figure*}
\includegraphics[scale=0.80]{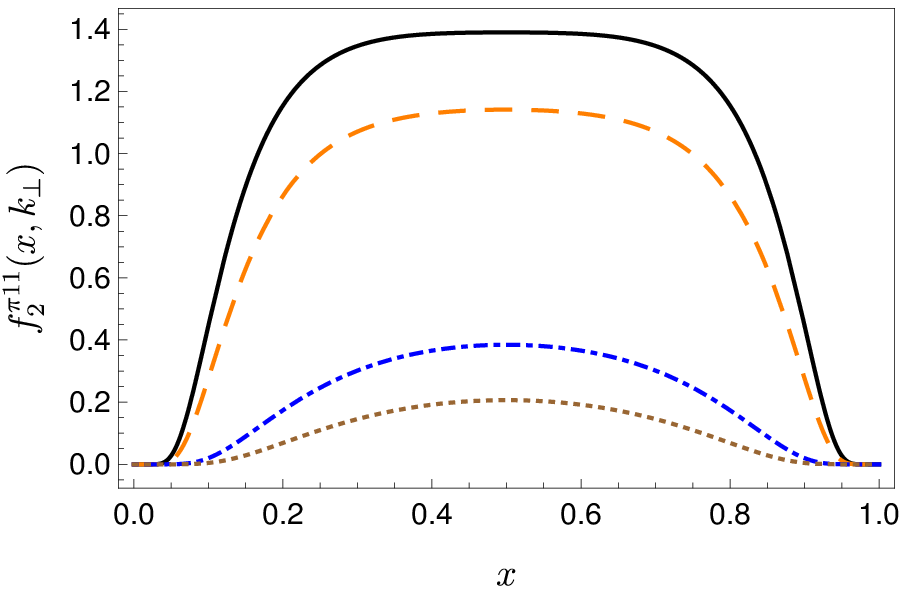}
\hskip 1cm \includegraphics[scale=0.80]{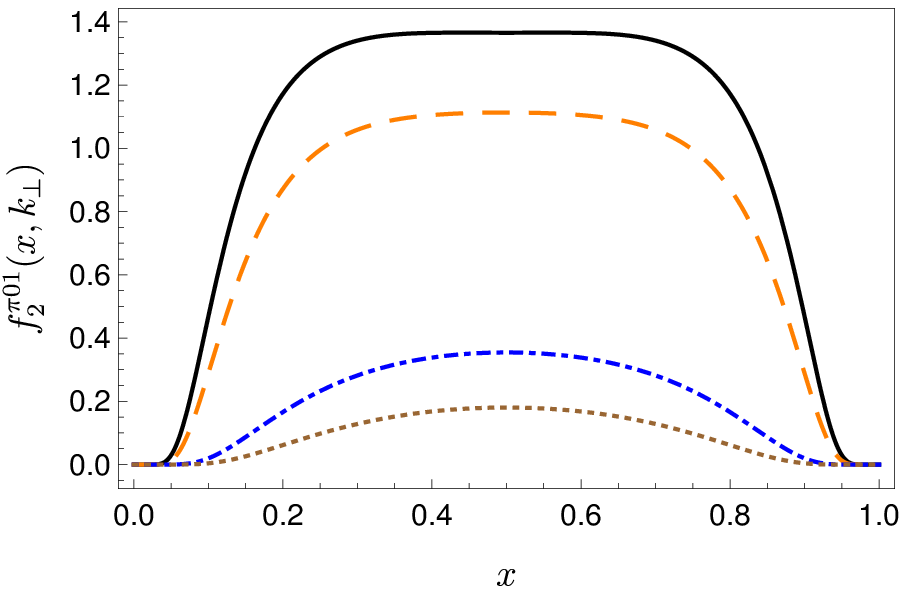}
\caption{\footnotesize  \textsl{Double PDFs of the pion evaluated 
within 
the 
dynamical spin effects model
\cite{spin} 
for 
different values of $k_\perp$. Full black line $k_\perp=0$ GeV, dashed 
orange line 
$k_\perp=0.2$ GeV, dot-dashed blue line $k_\perp=0.5$ GeV and dotted 
brown line 
$k_\perp 
= 0.6$ GeV. Left panel for the $A=B=1$ configuration. Right panel for 
the $A=0,~ B=1$ configuration. }}
\label{dpdfps}
\end{figure*}

\begin{figure*}
\includegraphics[scale=0.80]{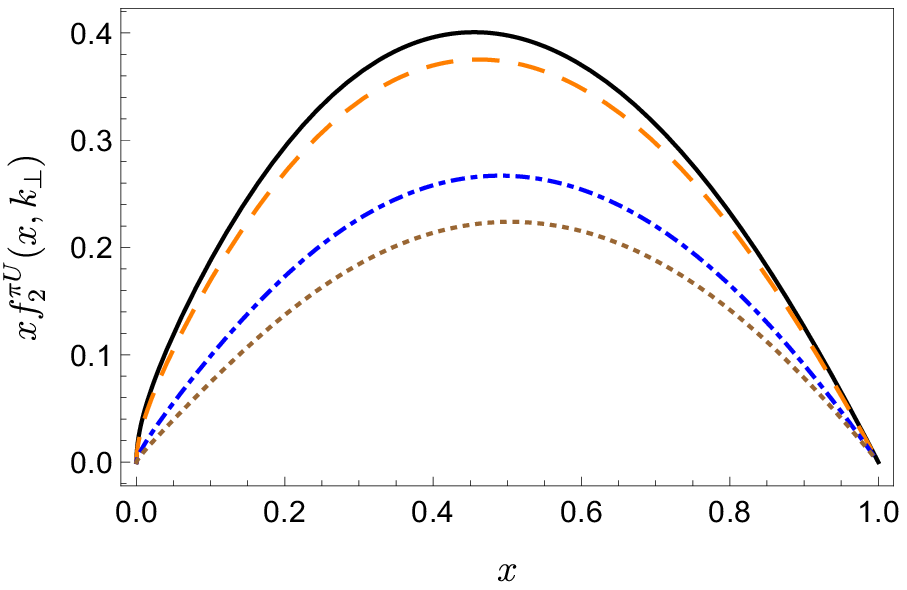}
\hskip 1cm \includegraphics[scale=0.80]{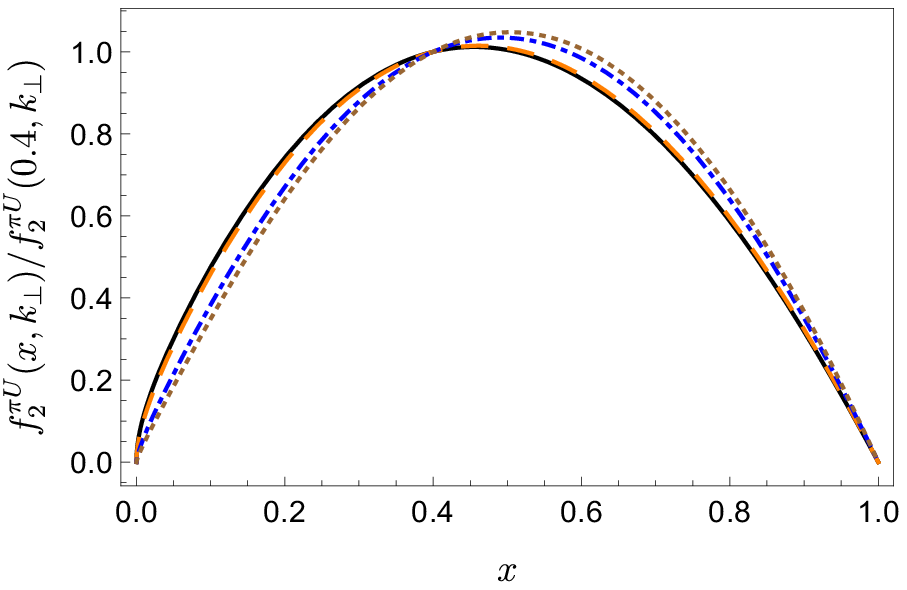}
\caption{\footnotesize  \textsl{Same of Fig. \ref{dpdfpo} but for the 
universal 
model of Ref. \cite{universal}. In the left panel, the quantity $x 
f_2^{\pi 
U}(x, k_\perp)$ is plotted. }}
\label{dpdfU}
\end{figure*}

\begin{figure*}
\includegraphics[scale=0.80]{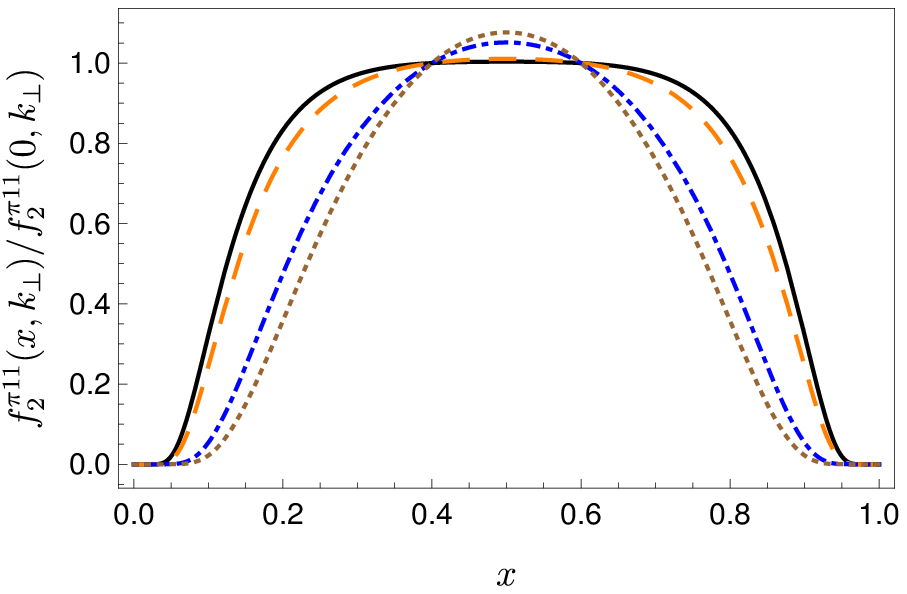}
\hskip 1cm \includegraphics[scale=0.80]{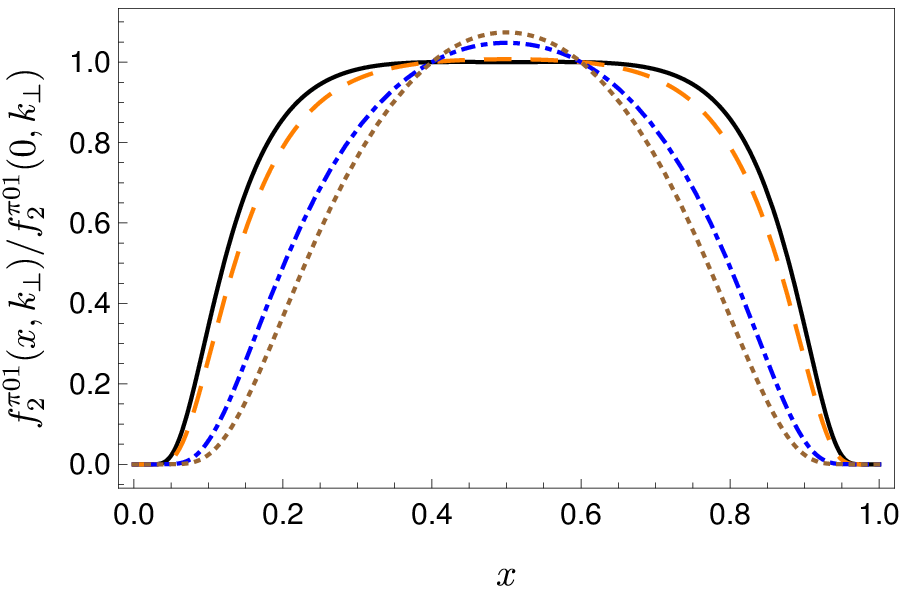}
\caption{\footnotesize  \textsl{The ratio $r_k(x,k_\perp)$, Eq. 
(\ref{rk}) 
evaluated for for different values of $k_\perp$. Full black line 
$k_\perp=0$ 
GeV, 
dashed orange line 
$k_\perp=0.2$ GeV, dot-dashed blue line $k_\perp=0.5$ GeV and dotted 
brown line 
$k_\perp 
= 0.6$ GeV.  Left panel for the $A=B=1$ configuration. Right panel for 
the $A=0,~ B=1$ configuration.} }
\label{factp_s}
\end{figure*}

In the left panel of  Fig. \ref{dpdfpo}, Fig. \ref{dpdfps} and  the left 
panel of  
Fig. \ref{dpdfU},
we show the calculations of the pion dPDFs~ (\ref{f2m}) for fixed 
different values of $k_\perp$.  In the 
cases 
of the original and dynamical spin models, the shape of these 
quantities are symmetric, reflecting the symmetry between $x$ and 
$1-x$, 
 see left panel of Fig. 
\ref{dpdfpo} and Fig. \ref{dpdfps}. The different behaviour, observed 
in the case of the  universal model,  is 
related to the  implementation of the Regge trajectory at small $x$.  
A common 
feature shared by all these models is the decreasing shape w.r.t. the 
increasing of 
$k_\perp$. Such a result is directly related to the behaviour 
of 
the relative eff. Details on the 
evaluation of 
the latter quantity are presented later on this section. 
As shown for the proton case~ \cite{noi1,noij2}, the impact of DPCs 
effects is 
enhanced for dPDFs depending  on $x_1-x_2$ with $x_1$ and 
$x_2$ which are almost independent and bound by $x_1+x_2\leq 1$. 
However, for a meson, where only the two body Fock 
state contribution is considered in Eq. (\ref{state}), the dPDF  
depends only on $x_1=x$ and $x_2 = 1-x$ due to momentum conservation~ 
\cite{noipion}.  For the moment being, a full expression for the LF 
wave 
function corresponding to, e.g.  
 a $|q \bar q q \bar q \rangle$, is not 
available. In fact, let us remind that, in 
the UM, such a contribution is included only to describe
 PDFs, thus a possible non trivial dependence of the dPDF on 
$x_1,~x_2$ and $x_3$ is not addressed.
In this scenario,
the most relevant sign of DPCs is given by studying the $x-k_\perp$ 
dependence of dPDFs.
In particular, an unfactorized dependence of dPDFs, w.r.t. the $x$ and 
$k_\perp$, 
 represents 
a possible signals of   double parton correlations.
We recall here that in order estimate the impact of these effects, the 
ratio $f_2^\pi(x,k_\perp)/f_2^\pi(x=0.4,k_\perp)$ is evaluated as a 
function of $x$ for different values of $k_\perp$. One should notice 
that if correlations were neglected, then the latter quantity would 
be 
constant w.r.t.  variations of $k_\perp$.
As one can see in the right panel of Fig. \ref{dpdfpo} and in Fig. 
\ref{factp_s},  correlations are very strong for the original and 
dynamical spin models. However, 
as one can observe in the right panel of Fig. \ref{dpdfU}, the impact 
of DPCs, encoded in the UM, is less relevant w.r.t. the other models. 
This feature is related to the poor general knowledge of these effects.
In any case, for all models here considered, 
 the factorisation in 
the 
$k_\perp$ and $x$ 
dependence is not fully supported.

\subsection{The pion form factor}
\begin{figure}
\includegraphics[scale=0.8]{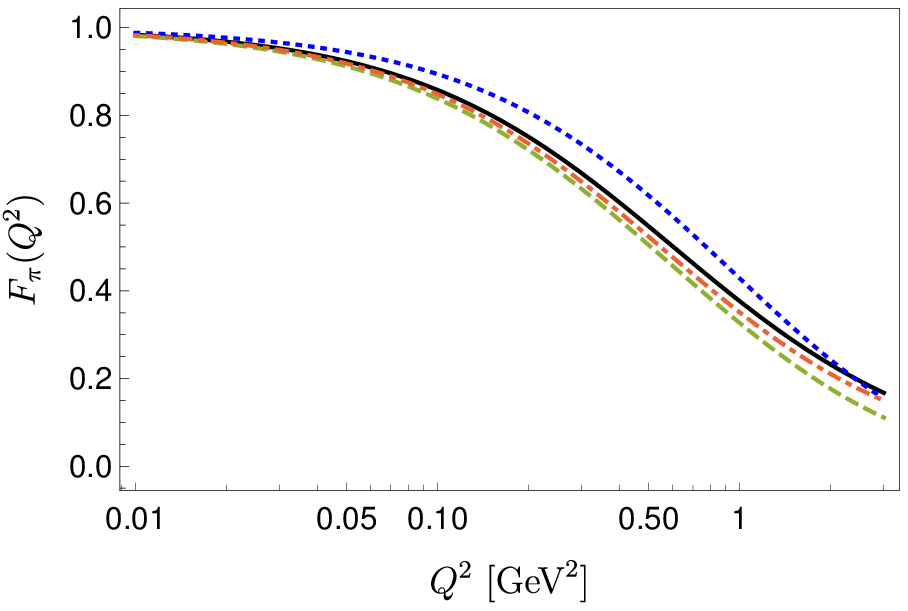} \hskip 0.3cm
\includegraphics[scale=0.80]{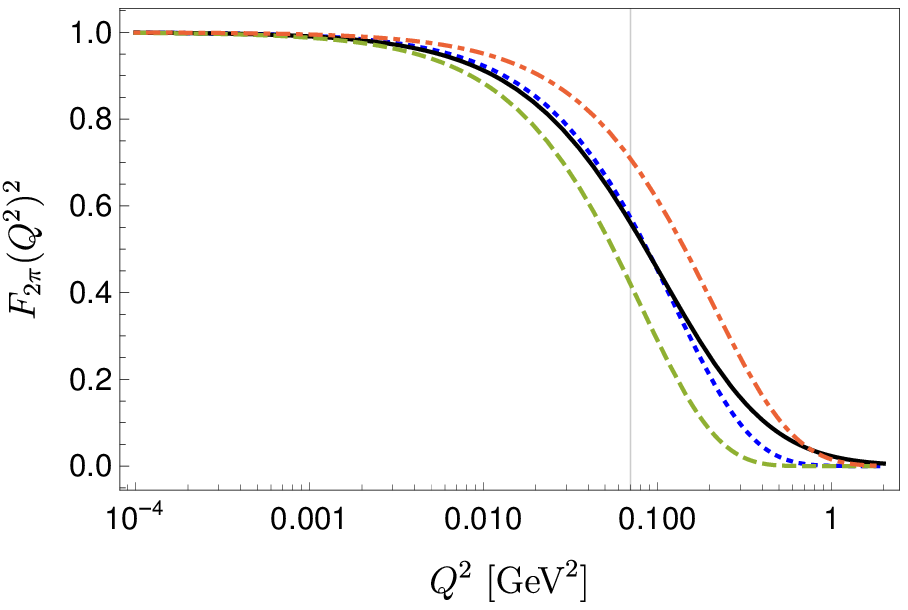}
\caption{\footnotesize  \textsl{ Left panel: the pion form factor 
evaluated within the lattice framework (full black
line),
the original model 
(dotted blue line), the dynamical spin model (dashed green line), in 
the $A=B=1$ 
configuration, 
 and the universal model (dot-dashed red line). Right panel, same of 
the 
left panel for the square of 
the eff 
$F_{2 \pi}(Q^2)^2$.
}}
\label{pionff}
\end{figure}

Since the main purpose of the present study is to compare 
 lattice data  with holographic
quark model 
calculations, here we show results for the pion e.m.
form factor. This quantity has been extensively 
investigated from a theoretical and experimental point of view 
\cite{lattice,Br1,universal,spin,seth}.
To this aim, we consider  the
pion ff evaluated within the lattice techniques in the $A$ 
configuration.
As one can see in the left panel of Fig. \ref{pionff}, the AdS/QCD 
approach is able to reproduce the essential behaviour of the pion ff.
In particular, the original model \cite{Br1} fits the ff in the small
 $Q^2$ region,
while the dynamical spin and universal ones
 \cite{universal,spin} 
provide an impressive agreement.
However,  
 by comparing the values of the mean pion radius,
one can conclude that the model which includes dynamical spin effects
reproduce very well experimental data \cite{pdg}, 
 see Table
\ref{tabr}.

\begin{table}
        \centering
        \label{1}
        \begin{tabular}{|c|c|c|c|c|c|c|}
            \hline
             & Original  & Dynamical Spin&Universal&Lattice&Lattice&
Experiment\\
& model &$A=B=1$&  model& (A) & (B) & \cite{pdg}\\
            \hline  
            $\sqrt{\langle r^2 \rangle}$ [fm] & 0.524&0.673 & 
0.644 &0.600&0.621&0.67 $\pm 0.01$ \\
            \hline          
        \end{tabular}
        \caption{ \footnotesize Values of the  pion mean radius, Eq. 
(\ref{ratio1}), 
obtained within the lattice and the models based on the the AdS/QCD 
approach. Experimental data are from Ref. \cite{pdg}. }
\label{tabr}
\end{table}

\subsection{The effective form factor of the pion}
Here we show the first comparison between the calculations of the eff 
within AdS/QCD inspired models and that from lattice QCD, see Eq. 
(\ref{latticeff2}).
Let us first discuss some differences between the pion ff and  eff.
As discussed in Refs. \cite{noij3,noiplb}, in the proton case, the two 
objects are completely different. In particular the eff involves two 
particle correlations and depends on  $k_\perp$, i.e. the momentum 
unbalance between the first 
and the second parton in the initial and final states.
In the e.m. form factor, $q_\perp$  represents the 
exchanged momentum 
between the initial and final state of a given parton.
 However, in the 
mesonic case, if one considers only the $|q \bar q \rangle$  
contributions, the 
formal 
expression of the eff (\ref{eff2})  and the e.m. (\ref{ffc}) one are 
extremely similar (see Ref. \cite{noipion} 
for details on this topic).
In addition,  $k_\perp$   
represents the conjugate variable to $d_\perp$, i.e. the transverse 
distance between the two partons, while $q_\perp$ is the conjugate 
variable to $r_\perp$, i.e. the transverse distance of a parton w.r.t. 
the centre of the hadron. 
In the right panel of Fig. \ref{pionff}, results of the calculations of 
$F_{2 \pi}(Q^2)^2$ has been shown. 
We remind that
this combination of two effs 
 enters the expression of $\mse$, see Eq. (\ref{sieffa}), and
encodes the hadron geometrical properties  which affect this 
experimental observable. 
Thus, in the 
right panel of Fig. \ref{pionff},
we 
highlighted the main discrepancies, between lattice and model 
calculations,
which also  affect the mean value of $\se$.
 As one can see, only the original model
 is able to 
reproduce the eff in the allowed kinematic region.
Let us stress again that the comparison is well motivated only for $Q^2 
< 
m_\pi^2$. In the forward region, frame dependent effects are important 
but not 
included in the LF formalism. 
 It is 
fundamental to point out that, while in the lattice framework there are 
no 
truncation of the meson Fock state, in all AdS/QCD models, but the UM 
one, only the first $|q \bar q \rangle$ contribution is included. Thus 
the dPDF is restricted to be considered as an unintegrated PDF where 
the
momentum conservation unambiguously   fixes the relation between $x_1$ 
and $x_2=1-x_1$.   In this scenario, lattice data of the first  
moment 
of 
dPDFs 
of the pion represent a reach starting point to understand in details 
the contribution of high Fock states in the  meson expansion Eq. 
(\ref{state}). Thanks to 
this analysis, 
further  implementations of holographic models could include two-body 
effects based on the  lattice data \cite{lattice}. As one can see in 
Table
\ref{tabd}, the lattice calculation of $\langle d^2 \rangle$ is 
comparable
to that obtained within the original model. On the contrary, the UM  
largely underestimates the mean partonic distance. 

\begin{table}
        \centering
        \label{1}
        \begin{tabular}{|c|c|c|c|c|}
            \hline
             & Original  & Dynamical Spin&Universal&Lattice\\
& model &$A=B=1$&  model& (A)\\
            \hline  
            $\sqrt{\langle d^2 \rangle}$ [fm] & 0.968&1.207 & 
0.767 &1.046 $\pm 0.049$ \\
            \hline          
        \end{tabular}
        \caption{\footnotesize Values of the 3D mean partonic distance 
in 
the pion, Eq. 
(\ref{2dist}), 
obtained within the lattice and the model based on the the AdS/QCD 
approach. }
\label{tabd}
\end{table}

\subsection{Calculation of $\mse$}
Here we discuss a possible prediction for an ideal DPS process 
involving two pions. Due to the lack of data and experimental 
analyses, we focus on the mean value $\se$  
(\ref{sieffa}). In this scenario, 
 only  geometrical properties affecting $\se$ have been taken 
into 
account.
The evaluation of this quantity within the lattice framework would be 
extremely valuable in order to guide future experimental analyses.
However, as extensively discussed in the previous sections, lattice 
data have been obtained in the pion rest frame.
Therefore a direct phenomenological prediction cannot be safely 
obtained. 
However, one can
evaluate the
mean  value of 
$\se$ by changing the higher extreme value of the integral in Eq. 
(\ref{sieffa}).
 In fact, for 
$k_\perp < m_\pi$, frame dependent effects are small.
For the purpose of the present investigation, in Table \ref{tabse}, we 
have reported the results of the calculations of  
$\mse$. As one can 
observe  in the first row, up to $m_\pi$, the original model predicts 
a 
$\mse$ very close to that obtained from the lattice.
Let us remind that 
 the full value of $\mse$, evaluated within this model, 
 has been used in the experimental 
analysis of Ref. \cite{Koshkarev:2019crs}.
This result is 
completely coherent with the comparison between the eff evaluated 
within 
the lattice QCD and the original model.  From just a mathematical point 
of 
view, 
we also displayed the calculation of $\mse$ in the full range of 
$k_\perp$. As one can observe, 
the universal model provides a good fit with lattice results. 
However, let us 
stress again that in this case, frame dependent effects, preventing a 
clear 
comparison between lattice   and holographic calculations, 
cannot be neglected.
The full evaluation of $\se$ is anyhow relevant to verify the validity 
of the RC inequality (\ref{ine}).
As one can observe in Table \ref{tabine}, 
the latter perfectly works for all models and lattice calculations. Let 
us stress again that the relation 
between the mean value of $\se$ and the mean distance between two 
partons has been obtained in a complete general manner in Ref. 
\cite{rapid}. Therefore, the validation of the RC inequality, in model 
independent frameworks, such as the lattice QCD, is
extremely precious.

\begin{table}
        \centering
        \label{1}
        \begin{tabular}{|c|c|c|c|c|c|}
            \hline
             & Original  & Dynamical Spin&Dynamical Spin 
&Universal&Lattice\\
& model &$A=B=1$& $A=0,~B=1$& model & \\
            \hline  
            $\mse^{0 ~\mbox{to} ~m_\pi  }$ [mb] & 76.2& 89.4 &90.7 & 
67.3 &77.7 \\
            \hline 
            $\mse$ [mb] & 38.3& 60.9&62.6 & 22.2 &26 \\
            \hline          
        \end{tabular}
        \caption{\footnotesize  Values of $\mse$ obtained within 
different pion models and the lattice approach, 
by taking into account only geometrical effects. In the first row 
$\mse$ 
has been evaluated by performing the integral in Eq. (\ref{sieffa}) 
from $0$ to $k_\perp \sim m_\pi$. In the second row, the full 
calculation of $\mse$ has been performed. }
\label{tabse}
\end{table}

\begin{table}
        \centering
        \label{1}
        \begin{tabular}{|c|c|c|c|c|}
            \hline
              Model    
& $\sqrt{\dfrac{\mse}{ 
3\pi}\dfrac{3}{2}}$ & $\sqrt{\langle d^2 \rangle}$ 
&$\sqrt{\dfrac{\mse}{ 
\pi}\dfrac{3}{2}}$\\
  & [fm] & [fm] & [fm]\\
\hline
Original  & 0.781  &0.968 & 1.352\\
            \hline 
Dynamical Spin  & 0.980 &1.207 & 1.697\\
            \hline 
Universal  & 0.594 &0.767 & 1.029\\
            \hline 
Lattice  & 0.647 &1.046 & 1.121\\
            \hline 
        \end{tabular}
        \caption{\footnotesize Check of the validity of the RC 
inequality (\ref{ine}) 
in 3-dimension. }
\label{tabine}
\end{table}


\subsection{Comparison between two-body distributions and the product 
of one-body 
functions}
\begin{figure*}
\includegraphics[scale=0.80]{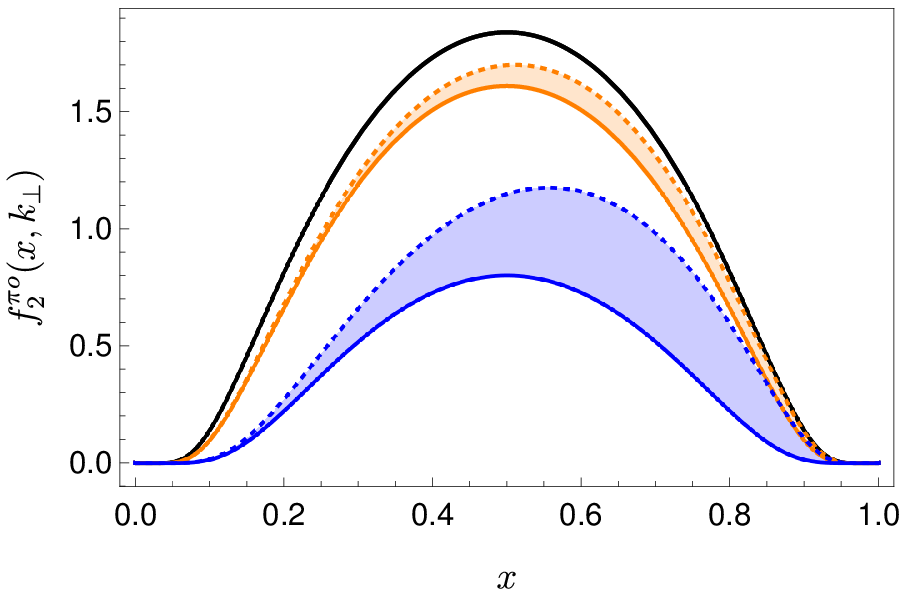} \hskip 0.5cm
\includegraphics[scale=0.80]{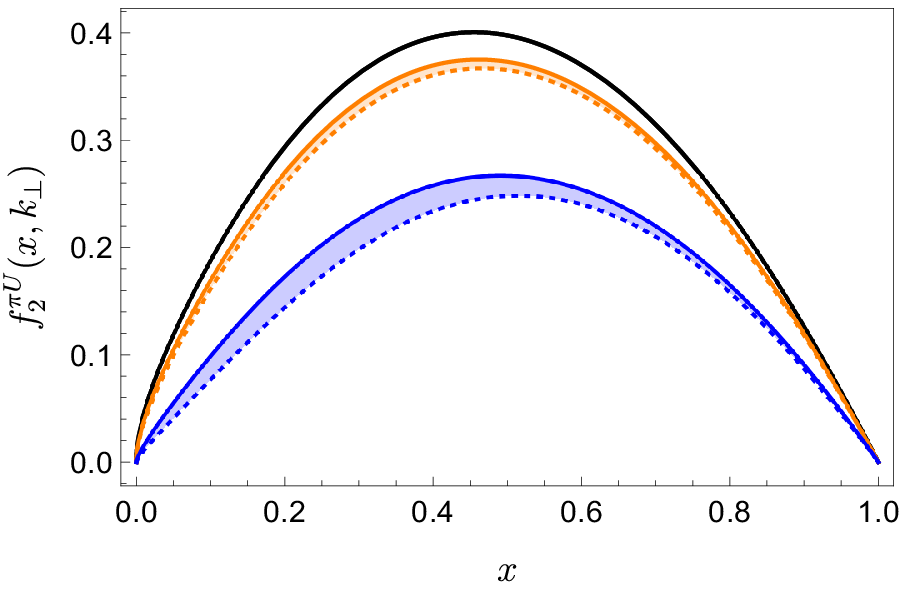}
\caption{\footnotesize  \textsl{
Full lines represent the pion dPDF
 Eq. (\ref{f2m}) and dotted lines stand for the
approximation Eq. (\ref{f2ma}). Quantities are evaluated for three 
values
 of $k_\perp$: 
$k_\perp=0$ GeV black lines, $k_\perp=0.2$ GeV orange lines 
and $k_\perp=0.5$ GeV 
blue lines.
The bands stand for the difference between the full calculation of the
 dPDF and 
its approximation.
Left panel 
for the original model of Ref. \cite{Br1}. 
 Right panel for the  UV of Ref. 
\cite{universal}. }}
\label{App_o_u}
\end{figure*}

\begin{figure*}
\includegraphics[scale=0.80]{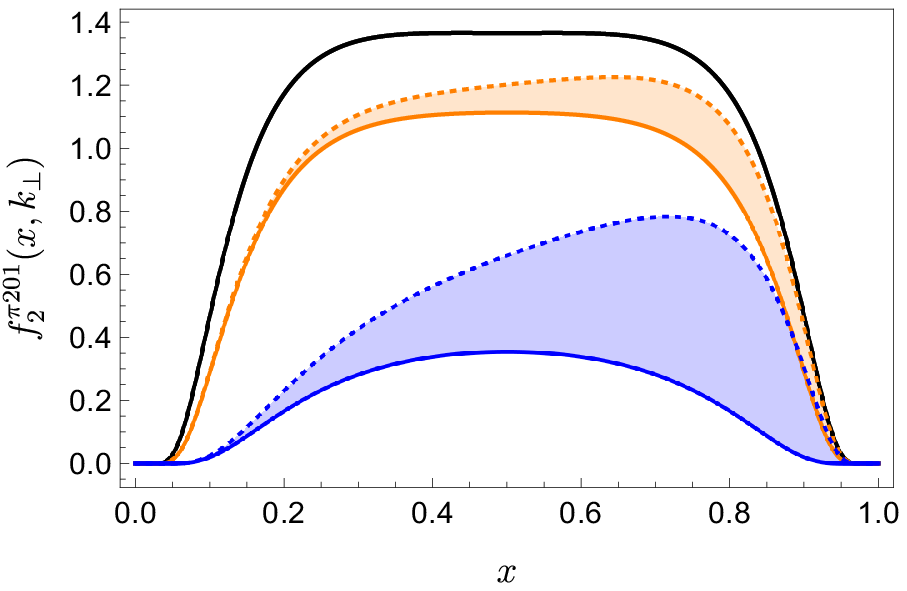} \hskip 0.5cm
\includegraphics[scale=0.80]{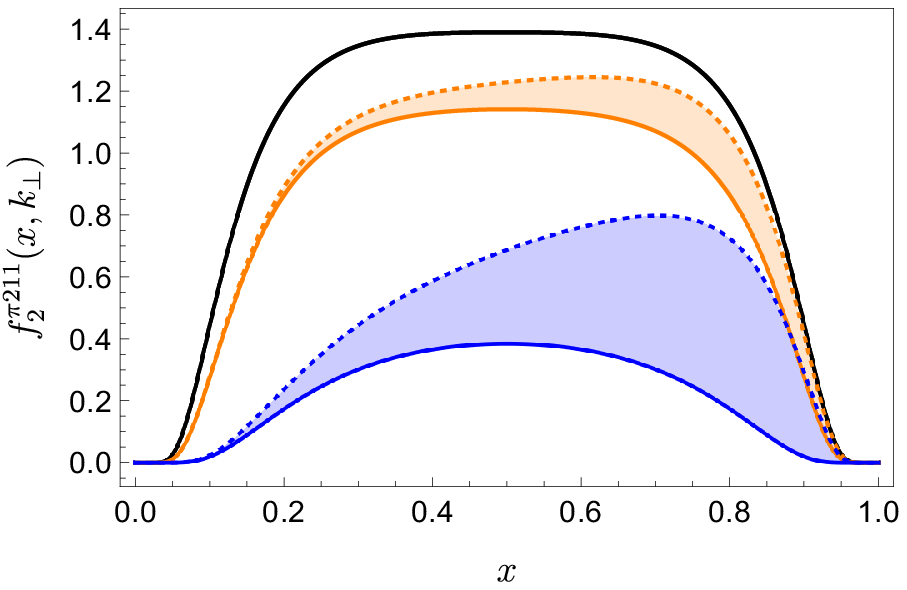}
\caption{\footnotesize  \textsl{
Same of Fig. \ref{App_o_u}  but for the dynamical spin model
 of Ref. \cite{spin}. Left panel 
for the configuration $A=0,~B=1$. Right panel:   for 
the configuration $A=B=1$. }}
\label{App_s}
\end{figure*}

\begin{figure*}
\centering
\includegraphics[scale=1.]{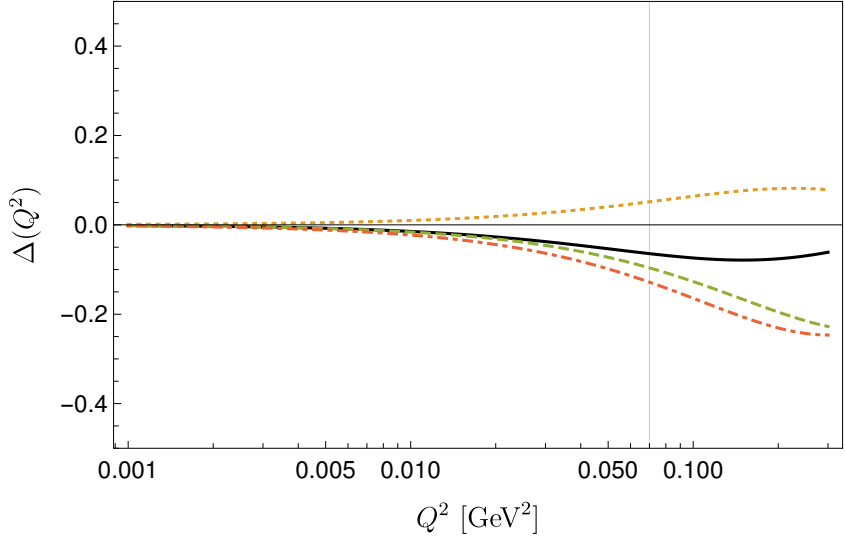} 
\caption{\footnotesize  \textsl{The quantity $\Delta(Q^2)$ 
(\ref{deltaeq}). Full black line stands for lattice data 
\cite{lattice}, 
dotted yellow line for the calculation performed within the universal 
model \cite{universal}, dashed green line represent the result of the 
evaluation within     the original model \cite{Br1} and 
dot-dashed red for the dynamical spin model \cite{spin} in the  $A=B=1$ 
configuration.  }}
\label{deltaF}
\end{figure*}

In this last part of this section, devoted to the study of the
 pion dPDFs, we 
discuss 
the validity of Eqs. (\ref{f2ma}) and (\ref{appmom}). Let us start with 
the comparison between
 $f_2^\pi (x, {\bf k_\perp})$
 and its approximation $f_{2,A}^\pi (x, 
{\bf k_\perp})$, i.e.
the 
product of the pion GPD and its form factor, see Eq. (\ref{f2ma}).
 Since the dPDFs of nucleons and 
mesons are basically unknown, 
in order to 
estimate the magnitude of DPS cross section, the approximation 
(\ref{f2c1a}) 
is often used. In this framework, model calculations
can be used to test
the validity of this ansatz.
Here we consider the same strategy developed in Ref. \cite{noipion}, 
i.e. we directly compare  $f_2^\pi (x, {\bf k_\perp})$ and 
$f_{2,A}^\pi (x, 
{\bf k_\perp})$ by remarking their differences. In 
 Figs. \ref{App_o_u} and \ref{App_s},  distributions have been 
evaluated for three different values of $k_\perp$  as functions of
 $x$.
As one can see, in all model calculations, but  the UM case, 
the shape of dPDFs is symmetric, at variance of the product of  ffs 
and GPDs. 
Such a feature can be explained by considering that
  GPDs and form factors depend on   the transverse 
momentum: ${\bf 
k_{1,\perp} } \pm (1-x){\bf k_\perp}$,
see 
Eq. (\ref{ffc}). Such a dependence, produces an 
asymmetry in the $x$ distribution, not  present in dPDF, see Eq. 
(\ref{f2m}). Moreover, since 
in the GPDs 
the momentum 
unbalance in the wave function is multiplied by the pre-factor
 $1-x<1$, the GPD goes to zero slower then the 
dPDF~\cite{noipion}. The 
last feature is partially
discussed 
in Ref. \cite{rapid} for the proton case. 
Furthermore, one should notice, that for the original and 
dynamical spin models the approximation Eq. (\ref{f2ma}) underestimates 
the full calculation of the dPDF, at the variance of  the universal 
case.  
In order to compare the impact of DPCs described within the lattice QCD 
and holographic models,
the following quantity will be also evaluated:

\begin{align}
 \Delta(Q^2)=F_{2\pi}(Q^2)-F_\pi(Q^2)^2~.
\label{deltaeq}
\end{align}
In order to minimise  frame dependent effects, we will focus on the 
region where $Q^2 < m_\pi^2$.
As one can see, if the approximation Eq. (\ref{appmom}) holds 
in some 
kinematic region, then the above quantity would be small. In   Fig. 
\ref{deltaF} 
 the calculations of the $\Delta(Q^2)$ function are displayed. As one 
can observe,
in the allowed region of $Q^2$, the original and dynamical spin models 
can almost 
reproduce the behaviour of DPC effects. 
In any case, one should notice that there is no  model able to 
reproduce both the effective and the e.m. form factors at the 
same time 
with the same precision. In fact, both the dynamical spin and universal 
models, fit very well data on the e.m. form factor but fail in the 
description of the eff. On the contrary, the original model can 
qualitatively 
reproduce both the effective and e.m. form factors only in the small 
$Q^2$ region. 
The main 
outcome of this 
analysis is the evident need of the inclusion of more Fock states in 
the 
pion expansion  (\ref{state}) necessary to include all possible DPCs 
in order  to describe both the e.m. and effective form factors.
{ Let us stress that since the universal model effectively 
takes into account the 
$|q \bar q q \bar q \rangle$ state, it is suitable to deeply 
investigate the impact of non trivial DPCs in the pion. Further 
studies on the top of that, beyond the present analysis, are on going.  
  }

\newpage

\subsection{Parameter dependence of the results}
In this final section about the pion target,
a short discussion about the parameter
 dependence of the 
main results will be presented.
In particular  the form factor, the eff and the function $\Delta(Q^2)$ 
will be evaluated by changing the model parameters in 
order to improve the agreement with lattice data.

\subsubsection{The Original model}
\begin{figure}[h]
\centering
\includegraphics[scale=0.85]{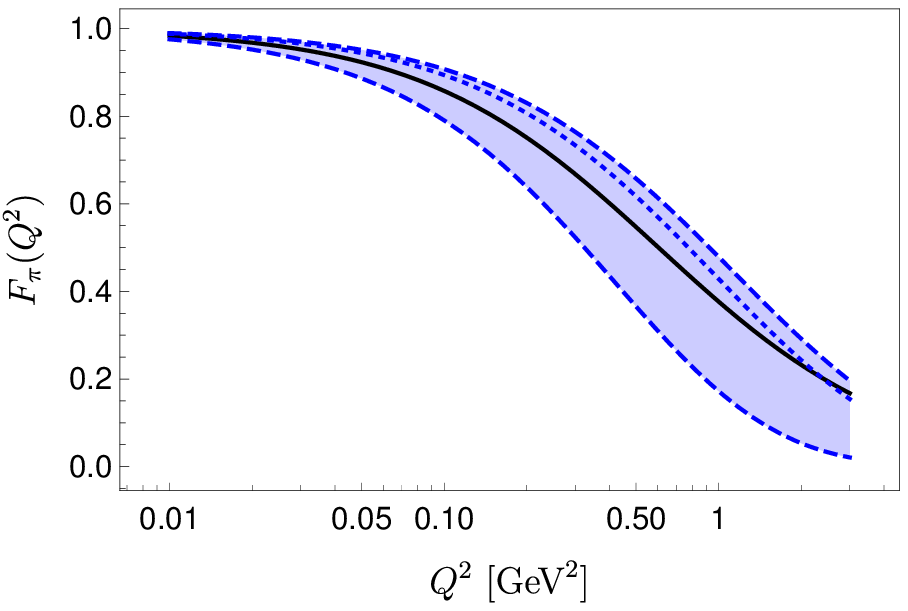} \hskip 0.3cm
\includegraphics[scale=0.85]{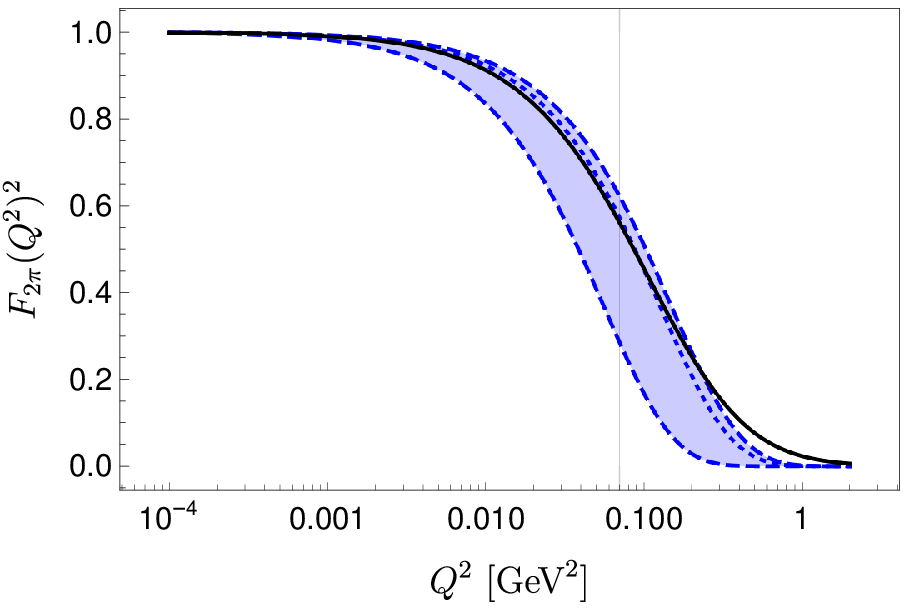}
\caption{\footnotesize  \textsl{Left panel: the pion form factor 
evaluated with the original model. The band stands for the results 
obtained by varying $\kappa$:  $ 0.35 
\leq \kappa \leq 0.59$ GeV. Dotted lines represents the result for 
$\kappa =0.548$ GeV. Right panel: same of the left panel for the 
quantity $F_{2\pi}(Q^2)^2$.
}}
\label{parorff}
\end{figure}

\begin{figure}[h]
\centering
\includegraphics[scale=0.85]{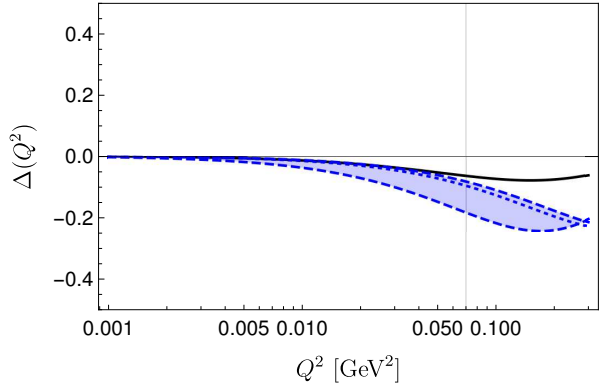} 
\caption{\footnotesize  \textsl{The function Eq. (\ref{deltaeq}) 
evaluated with the original model. The band stands for the results 
obtained by varying $\kappa$:  $ 0.35 
\leq \kappa \leq 0.59$ GeV. 
}}
\label{parordelta}
\end{figure}

As already mentioned, the the meson wave function evaluated within this 
model depends on two parameter $\kappa$ and $m_0$. Indeed, as shown in 
Ref. \cite{Brodsky:2014yha} the value of $\kappa$  smoothly depends on 
the observable one needs to fit. In fact, 
$\kappa$ obtained from the form factor is basically smaller then that 
obtained by fitting 
 the spectrum. In particular, 
$\kappa$ lies in
 the  range $ 0.35 \leq \kappa \leq 0.59$ GeV. 
Therefore here we present a selected collection of results, previously 
shown, as functions of $\kappa$. As one can see Fig. \ref{parorff}, the 
variation on $\kappa$ could allow to provide a good description of the 
form factors and the eff. Therefore, 
 a good agreement with lattice data could be obtained 
by adding a theoretical error on $\kappa$. One can interpret such an 
uncertainty as an attempt to include two-body effects in the wave 
function of the first Fock state.  In this scenario,  the mean radius 
and the main distance read: $\sqrt{\langle r^2 \rangle} = 0.625 \pm 
0.135$ fm and  
$\sqrt{\langle d^2 \rangle} = 1.17 \pm 0.28$ fm, respectively. The 
lattice and experimental results are included into the found ranges. 
Moreover, as one can see in Fig. \ref{parordelta}, the double parton 
correlations, emphasised by the function (\ref{deltaeq}), are well 
reproduced. However
the changes in $\kappa$ will produce modifications of the PDFs  and a 
different description of the spectrum.

\subsubsection{The Universal model}
\label{unmodel}
\begin{figure}[t]
\centering
\includegraphics[scale=0.85]{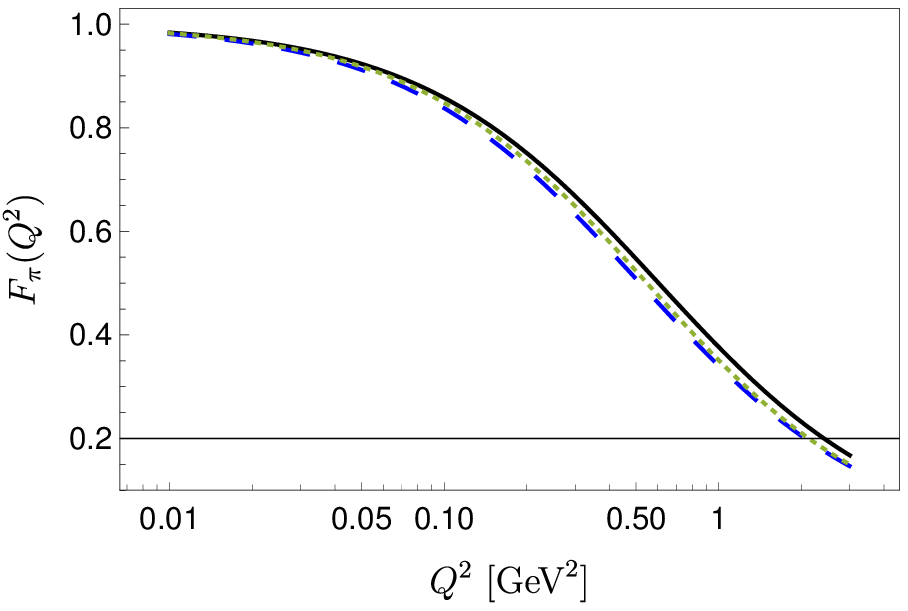} \hskip 0.3cm
\includegraphics[scale=0.85]{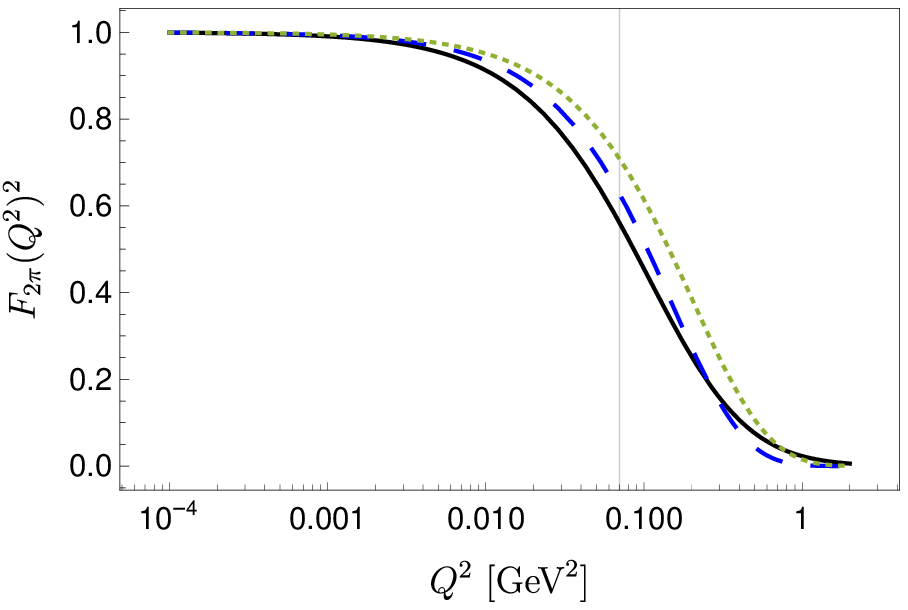}
\caption{\footnotesize  \textsl{Left panel: the pion form factor 
evaluated with the universal model. The blue dashed lines represent 
the calculation obtained by setting $a=1.4,~\sqrt{\lambda} = 0.51$ GeV 
and 
$\gamma = 0.01$. Dotted lines represents the result for 
the standard values of the parameters $a,~\lambda$ and $\gamma$, see 
Sect. \ref{unmodel}  for details.
 Right panel: same of the left panel for the 
quantity $F_{2\pi}(Q^2)^2$.
}}
\label{paruni1}
\end{figure}

\begin{figure}[t]
\centering
\includegraphics[scale=0.75]{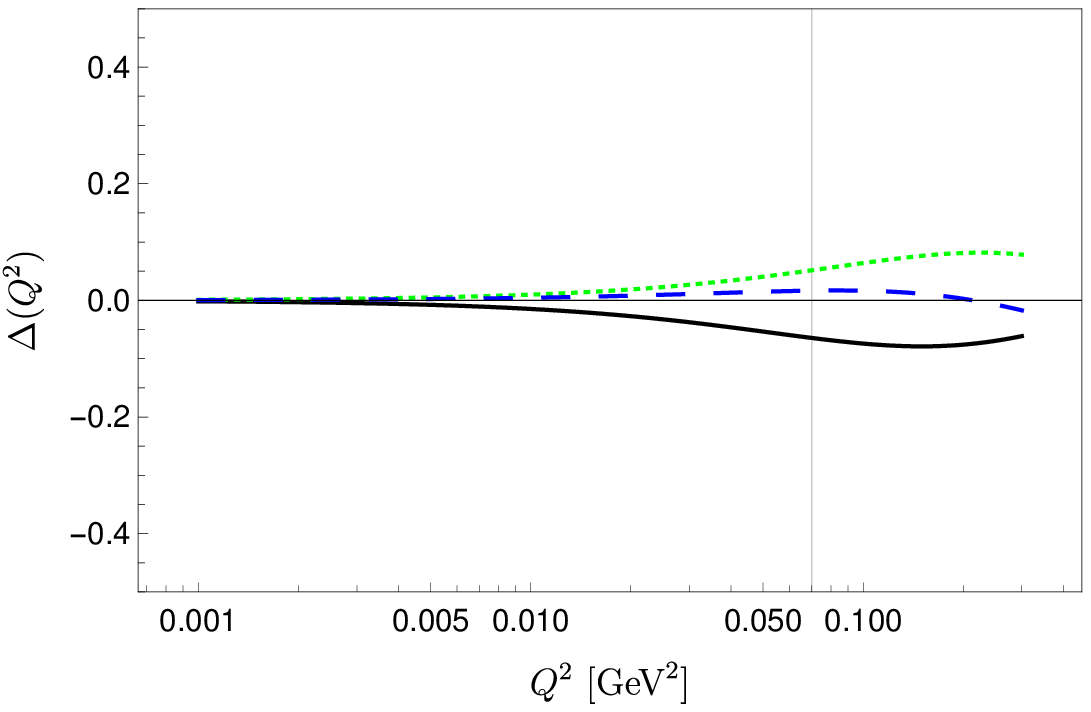} \hskip 0.3cm
\includegraphics[scale=0.85]{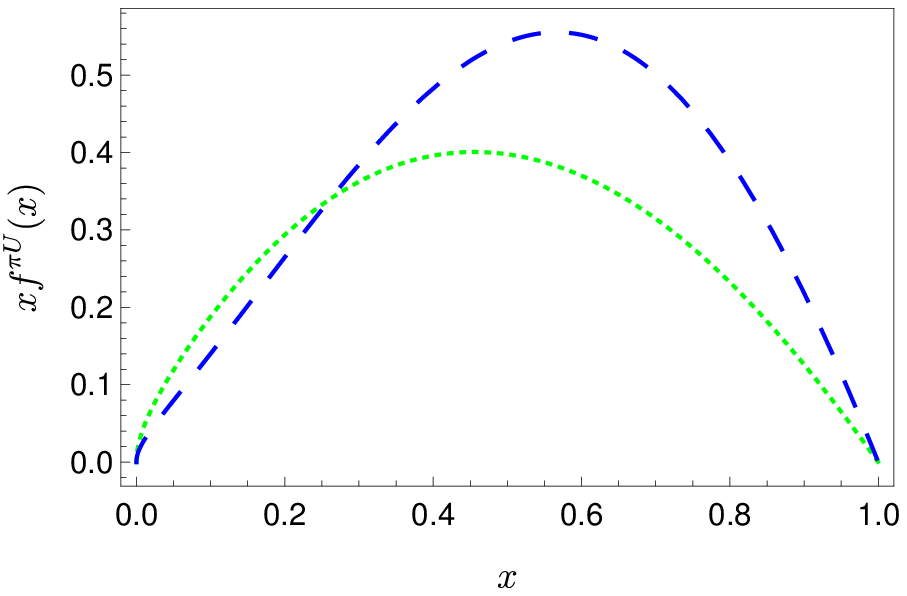} 
\caption{\footnotesize  \textsl{ Same of Fig. \ref{paruni1} but
left panel: for the $\Delta(Q^2)$ function. Right panel:  for the pion 
PDF.
}}
\label{paruni2}
\end{figure}

In the case of this model, use has been made of different parameters in 
order to properly  fit several observable, i.e.  $a, \lambda$ and 
$\gamma$, 
see Eqs. (\ref{unt1},\ref{unt2}) and (\ref{unt3}).
 In Figs. \ref{paruni1} and \ref{paruni2}, the 
previous calculations of the form factor, eff, PDF and $\Delta$ will be 
compared with those obtained within another choice of the parameters, 
i.e. $a=1.4,~\sqrt{\lambda} = 0.51$ GeV and 
$\gamma = 0.01$. As one can see in Fig. \ref{paruni1}, within this 
combination, both the form factor and the eff qualitatively reproduce 
the 
lattice data.
The function $\Delta(Q^2)$ 
 is also  closer to that evaluated 
within the lattice QCD w.r.t. that shown in Fig. \ref{deltaF}.
 However, as 
one can see in the right panel of Fig. \ref{paruni2}, the price for  
 this choice of the parameters is a relevant change of the shape of the 
PDF. Therefore one might expect a relevant loss of agreement between
PDF data and model calculations; in particular in the high $x$ region.
Since one of the motivation for the choice of original values of  $a, 
\lambda$ and $\gamma$ is also the excellent  fit with 
PDF \cite{universal}, 
one might conclude that a good strategy to explain lattice data, on 
DPCs, 
would be a further study on higher Fock states in the meson expansion.
In this scenario, the universal model is potentially very promising 
being the only one which effectively includes the $\bar q q q\bar q$ 
contribution. A detailed investigation on the LF wave function 
$\psi_{\bar q q q\bar q}$ guided by these lattice data would open a new 
window on the meson structure.

\newpage

\section{The $\rho$ meson within AdS/QCD }
\label{VI}
In this section, we introduce and calculate the dPDFs of the $\rho$ 
meson.  This is the first analysis of DPS involving a vector meson. 
However, in the future moments of the $\rho$ dPDFs
 could 
be accessed via 
lattice techniques. Here we consider 
possible predictions provided by AdS/QCD based models.
In order to evaluate the $\rho$ w.f., the procedure developed in Refs. 
\cite{rho,fore2,fore3} has been adopted.
In particular, for the three polarisation of the $\rho$ meson, the 
wave 
function 
is built from that of the pion. In this case, the input will be the 
w.f. Eq. (\ref{pionwfori}).  In this case, the normalisation constant   
will 
depend on the $\rho$ polarisation. In Ref. \cite{fore3} the 
parameters have been chosen to describe several observable for both 
the $\rho$ and the $\phi$ mesons. In particular 
$\kappa_o = 0.54 $ GeV and $m_o=0.14$ GeV (configuration A). In 
the present work we propose and motivate  the use  of another 
combination  of  the 
parameters. In particular 
those used 
to calculate the pion w.f.  (\ref{pionwfori}), i.e. $\kappa_0 = 
0.548$ and $m_o=0.33$ GeV (configuration B), will be considered. A 
detailed analysis on this choice will be provided. Let us mention that 
 a good comparison with the 
moments of $\rho$ PDFs, evaluated within the lattice QCD, is obtained 
within the B configuration.
The $\rho$ w.f., built from that of the pion, reads as follows:

\begin{align}
\label{rhoL}
 &\Psi^L_{h,\bar h}(x,b_{\perp}) = {1 \over 2 } \left( 1+ { 
m_f^2- 
\bigtriangledown^2  \over m_\rho^2 x (1-x) }  \right) \phi_L(x, \zeta  
) \delta_{h, \bar h}&
\\
&\Psi^{T=\pm}_{h,\bar h}(x, b_\perp)= \pm \Big[ i e^{\pm i\theta} 
\Big(x 
\delta_{\pm h, \mp \bar h}  - (1-x)\delta_{\mp h, \pm \bar h}  
\Big)\partial_{b_\perp}
+m_f \delta_{\pm h, \pm \bar h}  \Big]{ \phi_T(x,\zeta) \over  2x(1-x) 
}~.&
\label{rhoT}
\end{align}
Where here and in the following, we denote with $\Psi^\lambda_{h,\bar 
h}(x,b_\perp)$ the 
$\rho$ 
meson wave function in  coordinates space for $\lambda=L,T$ 
polarisation 
and 
quark-antiquark helicities $h$ and $\bar h$, respectively. Moreover, 
$\zeta$ is 
the usual 
variable 
introduced in the AdS/QCD framework, i.e. $\zeta =  
\sqrt{x(1-x) }b_\perp  $. The symbol $\bigtriangledown^2 \equiv {1 
\over 
b_\perp } \partial_{b_\perp}+\partial^2_{b_\perp}  $.
The gaussian like function, appearing in 
Eqs.(\ref{rhoL}, \ref{rhoT}), describes the scalar part of the meson 
w.f. in 
AdS/QCD (\ref{pionwfori}):

\begin{align}
\label{bo}
 \phi_{\lambda}(x,\zeta) = N_\lambda \sqrt{x(1-x)} e^{-{\kappa_o^2 
\zeta^2 
\over 2 }  }e^{- { m_o^2 \over  2 \kappa_o^2 x (1-x) }  }~.
\end{align}

In Eq. (\ref{rhoT}), $b_\perp e^{i\theta}$ is the complex form of the 
vector 
${\bf b_\perp}$. 
 The normalisation 
condition of the $\rho$ w.f. is the following:

\begin{align}
 \sum_{h,\bar h} \int dx~ d^2{\bf b_\perp}~ \big| \Psi^\lambda_{h,\bar 
h}(x,b_\perp) \big|^2=1~. 
\end{align}

The normalisation constant, 
$N_\lambda$  appearing in Eq. (\ref{bo}), depends on the 
polarisation~\cite{rho}.
From the above expressions, the dPDF for a given polarisation 
 can be obtained as follows:

\begin{align}
\nonumber
 f_2^{\rho,\lambda}(x, k_\perp) &= \sum_{h,\bar h} 
\int d^2{\bf b_\perp}
 e^{i {\bf k_\perp} \cdot {\bf k_\perp} } \big| 
\Psi^\lambda_{h,\bar 
h}(x,b_\perp) \big|^2= 2\pi \sum_{h,\bar h} \int d b_\perp ~b_\perp 
J_0(b_\perp 
k_\perp)
\big| 
\Psi^\lambda_{h,\bar 
h}(x,b_\perp) \big|^2~.
\end{align}

In the transverse polarisation case, since dPDFs are diagonal 
distributions in 
coordinate space, the main quantity we need to evaluate is the following 
one:

\begin{align}
 \label{Twf}
&\sum_{h,\bar h}\big| \Psi^{T}_{h,\bar h}(x,\zeta)  \big|^2 ={ 
|\phi_T(x,\zeta)|^2 \over  4x^2(1-x)^2 }\Big[ 2m_f^2 +\kappa_\rho^4 
\zeta^2 x 
(1-x)(x^2+ (1-x)^2  )  \Big]~.
 \end{align}

 \subsection{Numerical results}
Here we show the numerical predictions for the dPDFs and effs of the 
$\rho$ mesons.
Since this is
the first analysis about this topic, the $\rho$ moments of PDFs have 
been used to
motivate the choice of the free parameters appearing in Eq. (\ref{bo}).
To this aim,
 we compare the calculations of these quantities, obtained within the 
AdS/QCD approach, 
with those addressed by the  lattice QCD~
  \cite{lattice2}. 
Furthermore,
also for this hadron, the role of DPCs in dPDFs and in effs will be 
investigated.
In addition, we have also 
calculated the mean value of $\se$
 in order to provide a first prediction  for
 this experimental quantity. Thus, for the 
moment being, only the geometrical 
contributions have been taken into account in the meson $\se$.

\subsection{The parameters entering the $\rho$ wave functions}
\begin{figure}[t]
\centering
\includegraphics[scale=0.75]{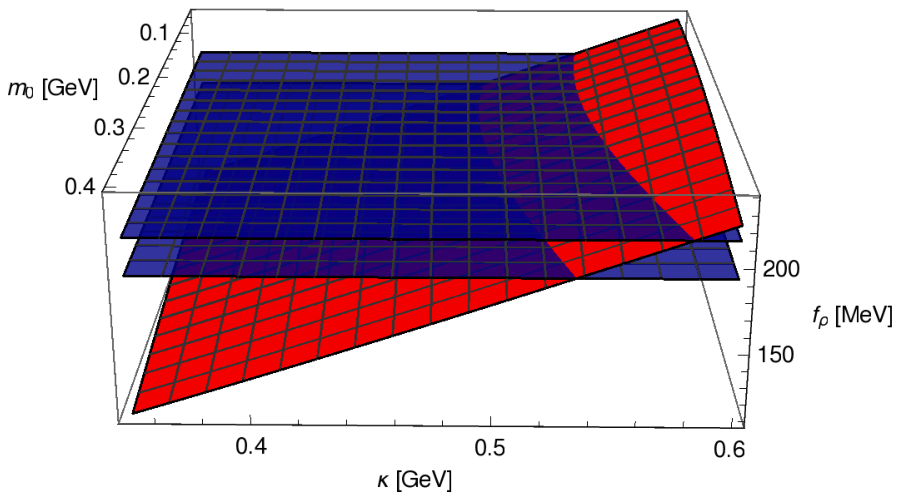} \hskip 0.3cm
\includegraphics[scale=0.75]{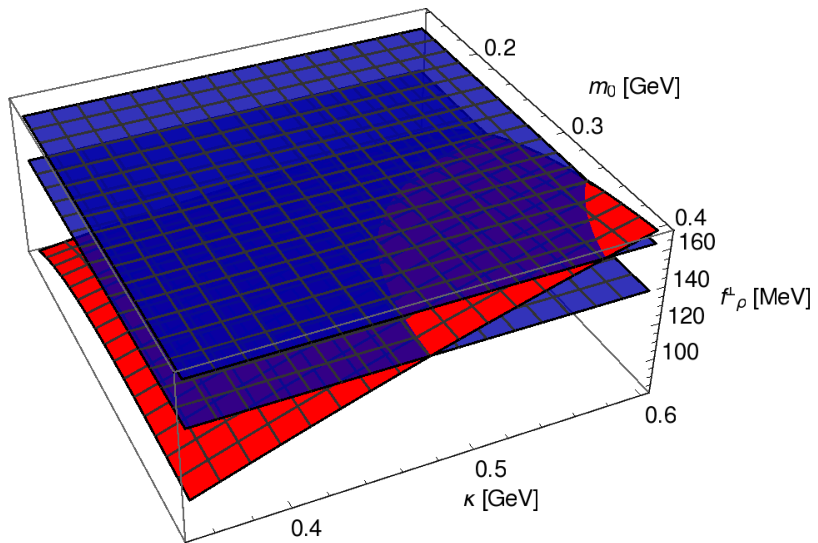}
\caption{\footnotesize  \textsl{ Left panel: the calculation of 
$f_\rho$ (\ref{fr1}) as a function of $m_0$ and $\kappa$ (the red 
diagonal plane). The blue horizontal planes represent the uncertainty 
obtained from sum rules \cite{sr1,sr2}. Right panel: same of the left 
panel but for $f_\perp^\rho$ (\ref{fr2}).
}}
\label{rhodecay}
\end{figure}

In Refs. \cite{rho,fore2,fore3} the parameters appearing in Eq. 
(\ref{bo}) have been properly chosen to reproduce the diffractive 
cross section for the $\rho$ and $\phi$ meson productions. Within the 
A configuration, also the decay constants are well reproduced. In the 
present 
analysis, we propose to use the parameters of the B configuration, in 
order to 
have a good agreement with lattice data of the $\rho$  moments of PDFs. 
To 
this 
aim,  we first show the different predictions for the decay constants, 
obtained 
within 
different combination of the two free parameters $\kappa$ and $m_0$.
We consider the following 
expressions:

\begin{align}
\label{fr1}
 f_\rho &= \sqrt{\dfrac{N_c}{\pi} } \int_0^1dx~ \left[1+ 
\dfrac{m_o^2- \bigtriangledown_b^2}{x(1-x)M_\rho^2} \right]\Psi^L(x,b) 
\Big|_{b=0}
\\
f_\rho^\perp &= \sqrt{\dfrac{N_c}{2 \pi}   }m_o \int_0^1 dx~\int db~\mu 
J_1(\mu b) \dfrac{\Psi^T(x,b)}{x(1-x)}~,
\label{fr2}
\end{align}
where $\Psi^\lambda(x,b)= \sum_{h, \bar h} \Psi^\lambda_{h, \bar 
h}(x,b)$, $b = |{\bf b}_\perp |$ and $\mu =1$ GeV. 
In Fig. \ref{rhodecay}, the above quantities are displayed as a 
function of the parameters $m_0$ and $\kappa$, respectively. As one can 
observe a good 
comparison with the sum rules  \cite{sr1,sr2} is obtained for a limited 
 choice of the parameters. Here below we compare in details the results 
obtained within the A and B 
configurations. In fact,  the former  leads to  good comparisons with 
data on diffractive $\rho$ production, while in the latter  
the 
same parameters entering the original model for the pion have been 
used. Let us 
recall that the w.f. of the original model represents the dynamical 
input used to evaluate 
the $\rho$ w.f., see Eq. (\ref{bo}). Therefore within the B 
configuration the $\rho$ eff can be calculated from the pion model which 
better reproduce the lattice data \cite{lattice} w.r.t. the other 
models. In 
addition let us mention that within the B configuration $\kappa = 0.548$ 
GeV, i.e. the same value adopted in both the original and universal 
pion models, thus reflecting the 
universal condition for  the breaking of the conformal 
symmetry.
As one can observe 
in Table \ref{tabrho2}, 
the results of the calculations of $f_\rho$ and $f_\rho^\perp$, within 
the A and B configurations, are very similar and comparable to those of 
 Refs. \cite{sr1,sr2}.

\begin{table}[h]  
\centering 
\begin{tabular}{llll} 
\hline\hline   
 Approach & Configuration  & $f_\rho$[MeV] & $f_\rho^\perp$ [MeV]  
\\ [0.5ex]  
\hline
 &  ~~~~~~A & 211 & 95 \\ [-1ex]
\raisebox{1.5ex}{LF holography} &   ~~~~~ B  
&204 & 150  \\[1ex]  
\hline 
Sum rules \cite{sr1}& & $198 \pm 7$ &$152 \pm 9$\\
\hline
Sum rules \cite{sr2}& & $206 \pm 7$ &$145 \pm 9$\\
\hline
\end{tabular}  
\caption{\footnotesize Predictions for the $\rho$ decay constants.} %
\label{tabrho2}
\end{table}

\subsubsection{Comparison with lattice QCD}
\begin{figure}[t]
\centering
\includegraphics[scale=0.75]{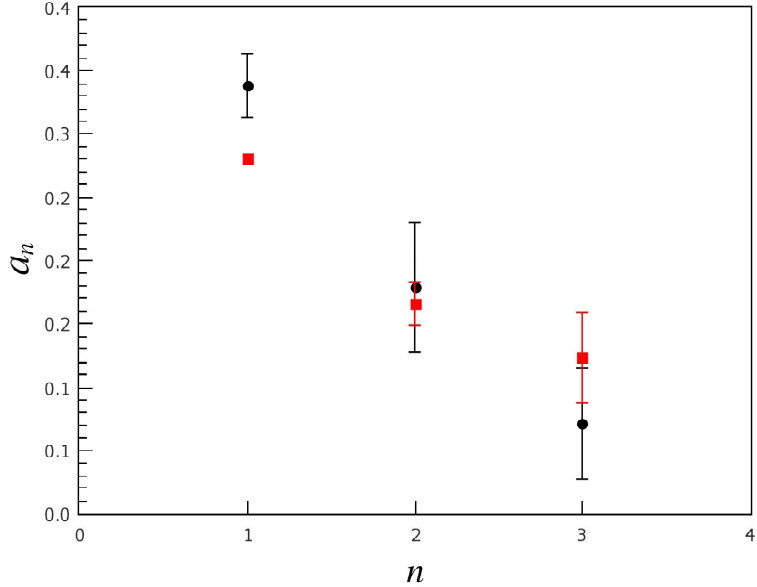}
\caption{\footnotesize  \textsl{Calculations of moments of the $\rho$ 
structure function Eq. (\ref{structure}). Round black points for 
lattice 
results. 
Squared red
points for model calculations obtained within the holographic model of 
Refs. \cite{rho,fore2,fore3} and including variations of $0.35 \leq 
\kappa \leq 0.6$ GeV.
}}
\label{latticerho}
\end{figure}

Here we discuss the comparison between 
the moments of the $\rho$ PDFs, evaluated within the holographic model~ 
\cite{rho,fore2,fore3}, with those obtained within the 
lattice QCD \cite{lattice2}. Let us remind that for the moment being 
their 
are no analyses 
of dPDF moments for the $\rho$ meson. Thus, in order  to investigate to 
what extent the adopted model could be 
compared to lattice predictions, here we only consider moments  of the 
following structure function:

\begin{align}
 \label{structure}
F_1(x)= \sum_q Q_q^2  \dfrac{1}{3} \Big(f_{1,q \bar q}^{\rho T}(x)+ 
f_{1,q \bar q}^{\rho L \uparrow }(x)  \Big)~,
\end{align}
  where here $q$ is the quark flavor with charge $Q_q$, $f_{1,q \bar 
q}^{\rho T}(x)$ is the $\rho$ PDF with transverse polarisation and 
$f_{1,q \bar q}^{\rho L 
\uparrow }(x)$ is the PDF of the $\rho$ meson longitudinally polarised 
and 
evaluated for a quark with positive helicity.  Since, in the 
holographic 
model, the 
latter quantity does not depends on the spin orientation, nor the 
flavor 
of the quarks, the above structure functions can be rewritten in terms 
of PDFs for unpolarised quarks:

\begin{align}
 \label{structure}
F_1(x)= \sum_q Q_q^2  \dfrac{1}{3} \left(f_1^{\rho T}(x)+ \dfrac{  
f_1^{\rho L}(x)}{2}  \right)~.
\end{align}
 In Ref. \cite{lattice2}, the following quantity has been calculated:

\begin{align}
 \label{momla}
a_n = 2 \int_0^1 dx~ x^{n-1}  F_1(x)~.
\end{align}
As one can observe in Fig. \ref{latticerho}, the first moment of the 
$\rho$ PDF is almost stable. 
Lattice data on $a_2$ and $a_3$ are contained inside the error bar 
which 
reflects variations of the model parameters.  Further 
improvements of the model are beyond the purpose of the present 
analysis. However the other moments are well reproduced and in 
particular in the B configuration one gets: $a_2 = 0.161$ and 
$a_3=0.102$ which are admitted by the theoretical error of lattice 
data. 
In 
further analyses,  implementations of the $\rho$ w.f. to improve the 
comparison with the 
lattice outcomes will be available. For example, one can 
use the pion wave function 
evaluated within the models of Refs. \cite{universal,spin}  as input of 
the procedure.

\subsection{Calculations of dPDFs, effs and $\mse$}
\begin{figure}[h]
\hskip-0.5cm\includegraphics[scale=0.85]{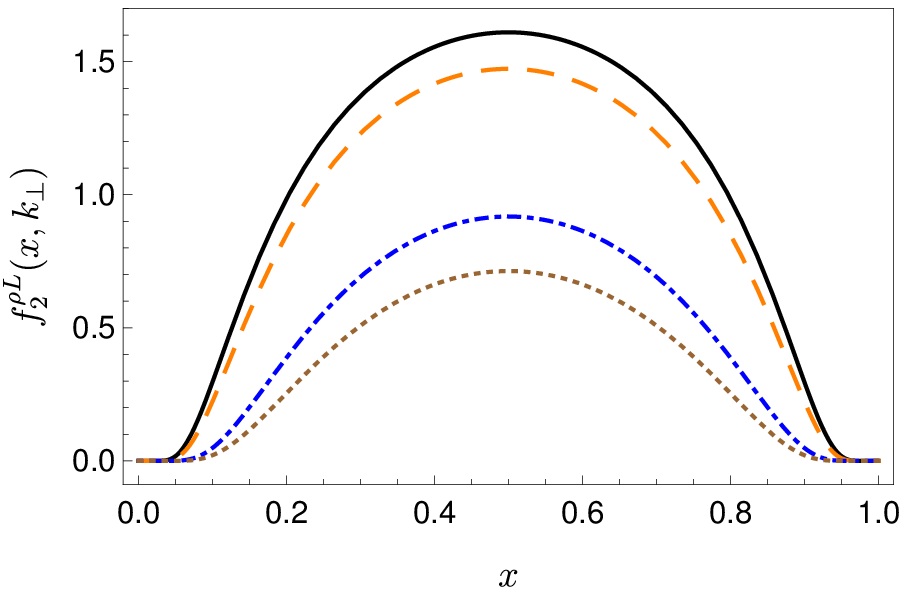}
\hskip 0.2cm\includegraphics[scale=0.85]{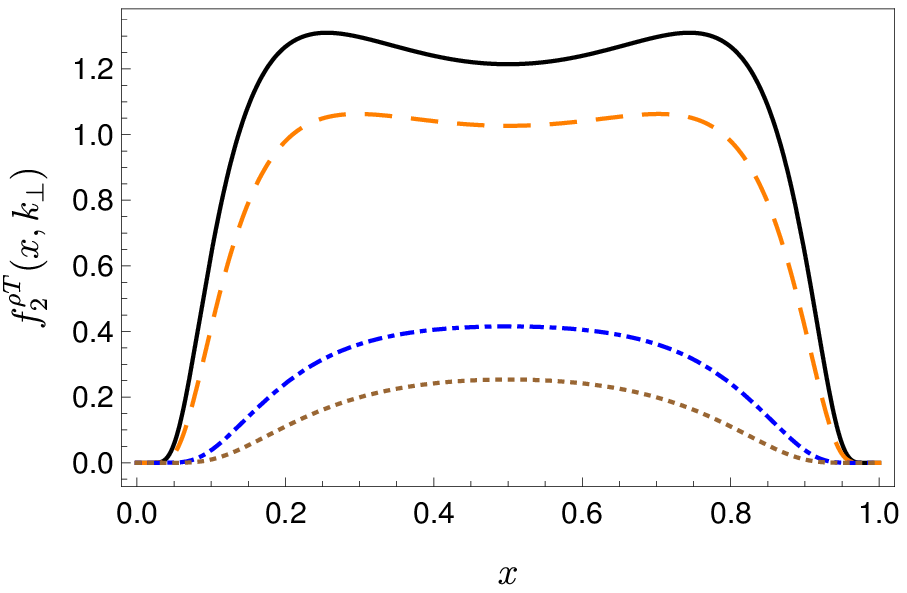}
\caption{\footnotesize  \textsl{
Same of Fig. \ref{dpdfps} for the
 $\rho$ meson. Left panel for longitudinal polarisation. Right panel 
for 
transverse polarisation.
}}
\label{factrL}
\end{figure}

\begin{figure}[h]
\hskip-0.5cm\includegraphics[scale=0.85]{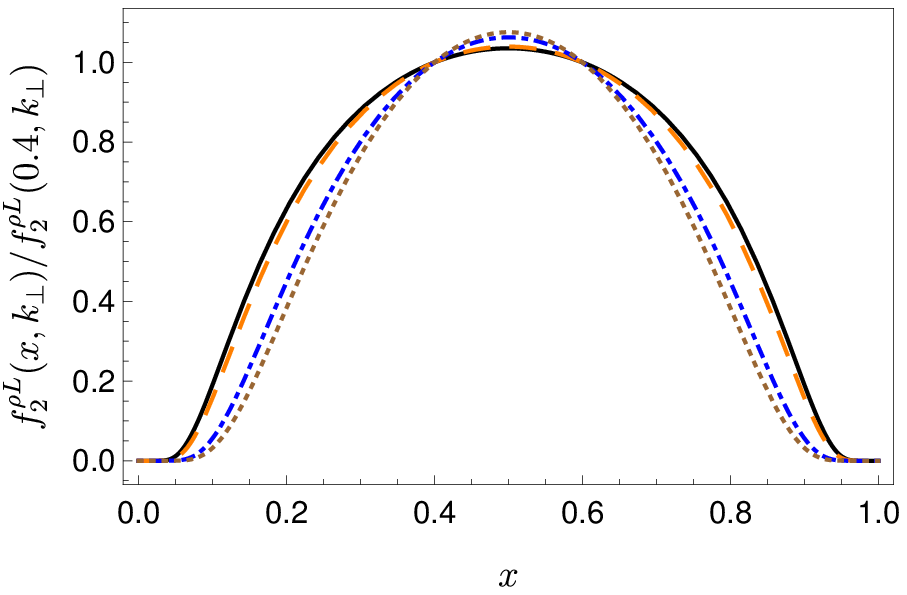}
\hskip0.2cm\includegraphics[scale=0.85]{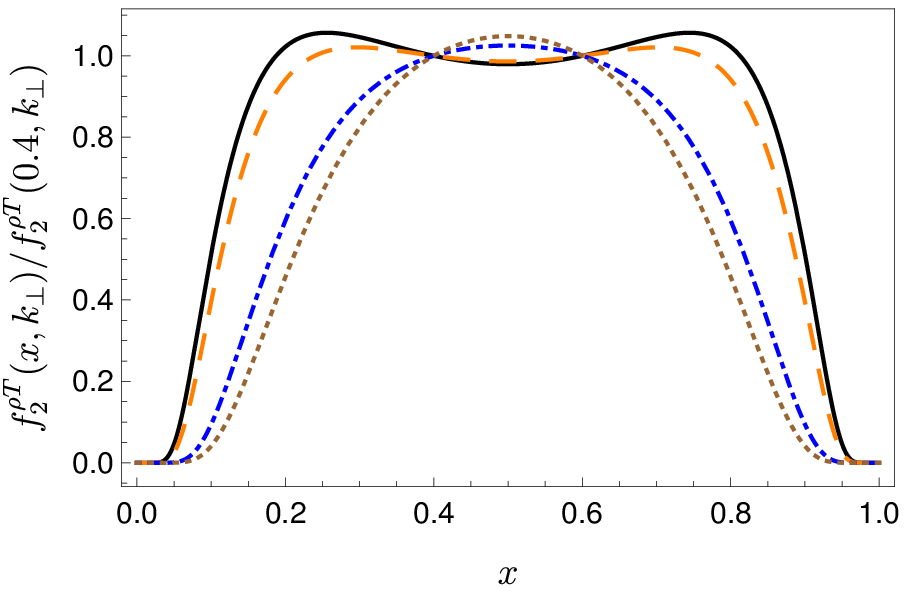}
\caption{\footnotesize  \textsl{
Same of Fig. \ref{factp_s} for the $\rho$ meson.
Left panel for longitudinal polarisation. Right panel for 
transverse polarisation.
}}
\label{factrT}
\end{figure}

\begin{figure}[t]
\centering
\includegraphics[scale=0.85]{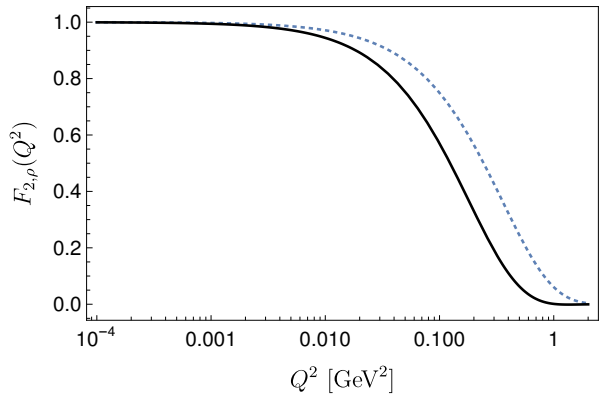}
\caption{\footnotesize  \textsl{The effective form factor Eq. 
(\ref{eff2}), 
 evaluated for the $\rho$
system. Full black line for the transverse polarisation and dotted blue 
line for 
the longitudinal one. 
}}
\label{f2rLT}
\end{figure}

In the present section, we show and discuss the results of  
numerical evaluations 
of $\rho$ dPDFs, effs and $\mse$ in the B configurations of 
the parameters entering Eq. (\ref{bo}). The above quantities have been 
calculated for the longitudinal and 
transverse 
polarisations separately.
As for the pion case, in  Fig. 
\ref{factrL}, we  display the $\rho$ dPDFs  for the 
two possible polarizations: left panel for the longitudinal 
polarisation and 
right panel for the transverse one, respectively. As one might notice, 
 the $k_\perp$ behaviour  of the distributions is similar to 
that 
obtained for the pion evaluated within the original models,  see left 
panel of 
 Fig. 
\ref{dpdfpo}. Such a 
result is coherent with the choice of the scalar w.f. entering  Eqs. 
(\ref{rhoL}) and (\ref{rhoT}).
Moreover,  in the transverse polarisation case, the distribution has 
two pronounced peaks.  
In addition, as one can see in Fig. \ref{factrT},  a 
possible factorisation between the $x$ and $k_\perp$ dependence is 
violated, thus reflecting the presence of correlations.
The amount of these effects is slightly different from those addressed 
in the 
right panel of Fig. \ref{dpdfpo} obtained within the original model. 
This feature is related to the presence of derivatives w.r.t. $b_\perp$ 
in Eqs. 
(\ref{rhoL}) and (\ref{rhoT}).
Thus, the overall dependence on $k_\perp$  of the  $\rho$ dPDFs 
is somehow different from that of the pion.
The main interpretation of the present outcome is that
 the procedure, used to  generate the $\rho$ w.f., 
 introduces additional correlations.
Moreover, the mean value 
of the effective cross section reads: 
$\overline{\sigma}_{eff}=27.8$ mb for the longitudinal case and
 $\overline{\sigma}_{eff}=54.7$ mb for the transverse one.
{ In order to provide a proper interpretation to these results, let 
us 
remark that the original model, used as input for the above 
calculations 
(\ref{rhoL}-\ref{rhoT}), qualitatively fits the Lattice data. Therefore 
it is reasonable to expect that also predictions for the $\rho$ could 
be 
realistic. Thereby one can conclude 
that the DPS 
cross section is dominated by the longitudinal component of the 
$\rho$ meson. In fact, we recall that the most $\se$ is small the most 
the DPS contribution is big with respect to the SPS case, see Eq. 
(\ref{pocket}).  In Fig. \ref{f2rLT} the eff of the $\rho$ meson, 
obtained by 
disentangling the two polarisation contributions, is shown. Full line 
represents the transverse polarisation and dotted line stands for the 
longitudinal one.  By using Eq. (\ref{2dist}), from the eff 
the mean partonic 
distance between two partons in the $\rho$ has been calculated: 
$\sqrt{\langle 
d^2 \rangle}=0.826$ fm and $\sqrt{\langle 
d^2 \rangle}=1.159$ fm, for the longitudinal and transversal 
polarizations, respectively.    
One should notice that in the former case 
the mean distance is lower then that evaluated for the pion target 
within 
the same original model \cite{Br1}, at the variance of the transversely 
polarised case,  see Table 
\ref{tabr}. 
Since, the original model predicts 
a mean value of $\langle d^2 \rangle$ in agreement to that of the  
the lattice QCD, one might expect
that valence quarks in the $\rho$ meson are closer 
to 
each other then  in the pion case if the the $\rho$ is 
longitudinally
polarised. 
This feature represents an extreme interesting prediction directly 
related to the non-perturbative structure of the $\rho$ meson.
Let us also show that 
the RC 
inequality (\ref{ine}) perfectly works also for the $\rho$ meson 
described 
within the
holographic approach, see Table \ref{tabiner}.
  }

\begin{table}
        \centering
        \label{1}
        \begin{tabular}{|c|c|c|c|c|}
            \hline
              $\rho$   
& $\sqrt{\dfrac{\mse}{ 
3\pi}\dfrac{3}{2}}$ & $\sqrt{\langle d^2 \rangle}$  
&$\sqrt{\dfrac{\mse}{ 
\pi}\dfrac{3}{2}}$\\
Meson  & [fm] & [fm] & [fm]\\
\hline
Longitudinal polarisation  & 0.665 &0.826 & 1.15\\
            \hline 
Transverse polarisation  & 0.933 &1.58 & 1.62\\
            \hline 
        \end{tabular}
        \caption{\footnotesize Check of the validity of the RC 
inequality (\ref{ine}) 
in 3-dimension for the $\rho$ meson. }
\label{tabiner}
\end{table}

\newpage

\newpage

\section{Conclusions}
Double parton distribution functions are new fundamental 
quantities encoding information on the three dimensional partonic 
structure of hadrons.
Double PDFs
enter the double parton scattering 
cross section for which   theoretical and experimental 
analyses are ongoing. However, for the 
moment 
being, only proton-proton and proton-ion collisions are  
investigated from an experimental  point of view. 
Nevertheless, lattice data on distributions related to the first moment 
of 
the pion dPDFs are 
now 
available. These quantities encode  double 
parton correlations which cannot be accessed via one-body functions 
such as standard form factors. This conclusion is qualitatively 
coherent with the quark model analyses for the proton target. The main 
purpose 
of the present study is to compare lattice QCD predictions of the 
effective form factor with  quark model calculations.
In particular, here we have considered 
AdS/QCD soft-wall inspired pion models for which
phenomenological implementations are also included.  Double PDFs have 
been 
calculated by 
showing their full dependence on the longitudinal momentum fraction and 
 the transverse momentum unbalance ${\bf 
k_\perp}$.  
Ratios sensitive to DPCs  have been calculated and results show that 
DPCs are relevant. An important comparison between dPDFs and 
their        
approximations in terms of  GPDs and form 
factors have been also investigated. Holographic model predictions 
shows 
that 
even if the pion is described by considering only the first $|q \bar q 
\rangle$ state, dPDFs cannot be described in terms of one-body 
functions. Such a conclusion is consistent with previous studies of the 
proton dPDFs. Let us stress here that such an approximation is largely 
used in phenomenological analyses of DPS processes.
In order then to provide useful predictions, an estimate of the 
experimental 
observable
$\se$ has been provided via quark models and lattice QCD. These 
results have been 
properly 
interpreted in terms of geometrical properties of the pion partonic 
structure by verifying the RC inequality. Furthermore, moments of 
dPDFs, i.e. the effective form factors, have been calculated within the 
adopted quark models and then compared with 
lattice data  for the first time. 
Despite the limited region in $Q^2$, which
minimises the impact of 
 frame dependent effects, one can 
conclude that for the moment being the absence of a complete evaluation 
of  high Fock states in the pion expansion  prevents a simultaneous 
description of the electric-magnetic and 
effective form factors.  Nevertheless, the 
original AdS/QCD model almost matches the lattice eff and qualitatively  
reproduces the impact of double parton correlations. On the 
contrary, even if the other models  provide an impressive description of 
the e.m. form factor, they 
fail in the evaluation of the eff. These first comparisons, between 
lattice and quark model analyses, point 
to 
the necessity of an accurate description of the contributions of high 
Fock 
states 
in the pion. The main conclusion is that lattice data can be used to 
add new constraints 
on  future implementations of holographic models. 
{ Let us mention that for the moment being, 
the only model which already  
effectively includes a  $q \bar q q \bar q$ contribution is the 
universal one. Therefore
the latter is very promising and suitable to describe both one-body 
quantities and DPCs in the meson at the same time; thus shedding a new 
light on the parton structure of the 
pion.    }
From another perspective, even if the frame dependence of the lattice 
eff prevents to get a phenomenological
  value of $\se$, the RC inequality has been 
inverted in order to provide a range of frame independent values of 
$\mse$ starting from the  
 $\langle d^2 \rangle$ addressed by  lattice QCD.  Such a procedure 
leads to a value of $\mse$, for a 
pion-pion collision, which is bigger then to the proton-proton case. 
This conclusion is directly obtained from lattice QCD data and could 
guide future experimental and theoretical analyses.
In the final 
part of this investigation, predictions for dPDFs and effs of the 
$\rho$ meson 
have been discussed for the first time. The main outcome of this 
analysis is that  
the impact of DPCs 
 change with the meson polarisation.

\section*{Acknowledgements}
MR acknowledges Sergio Scopetta,  Vicente Vento and 
Christian Zimmermann for precious
discussions.  This work was 
supported, in part by the 
STRONG-2020 project of the European Unions Horizon 2020 research and 
innovation programme under grant agreement No 824093, and by the 
project ``Photon initiated double parton scattering:
illuminating the proton parton structure'' on the FRB of the University 
of Perugia.

\appendix

\addcontentsline{toc}{section}{Appendix A}
\section*{Appendix A}
In this section  we discuss some details on the derivation of the dPDFs 
expression in 
terms of the LF wave function.
In particular, we make use
of the lattice conditions discussed in Sect. 3.
Let us remind that
the correlator matrix we need to evaluate is defined with quark field 
operators separated by a 
distance $y^\mu$ with $y^0=0$. Moreover, 
$\gamma^0$ is
the gamma matrix considered 
in the dPDF correlator. 
In this section we show in which kinematic conditions, frame dependent 
effects, due to the lattice conditions, are minimised.
To this aim, let us recall the main ingredients 
of the 
procedure. 
In particular, we have used the convention described in 
the 
Appendix A of Ref. \cite{Br1}. The quark field operators, defined in 
terms of 
light-cone coordinates, 
 reads:

\begin{align}
 q(x) = \sum_\lambda \int _{l^+>0}
 \dfrac{d l^+ d^2 \bf l_\perp}{\sqrt{2 
l^+} 
(2 \pi)^3}
~ \big[ b_\lambda(l) u_q(l,\lambda)e^{-i l \cdot x} 
+d_\lambda^\dagger(l) 
v_q (l, \lambda) e^{i l\cdot x}  \big],
\end{align}

where the anticommutation relation for the spinors reads:

\begin{align}
 \big\{b_\lambda(l), b^\dagger_{\lambda'} (l')    \big\} =  
\big\{d_\lambda(l), d^\dagger_{\lambda'} (l')    \big\}=(2 \pi)^3 
\delta(l^+-l'^+) \delta^{(2)}(\bf l_\perp-\bf l'_\perp ) 
\delta_{\lambda \lambda'}~. &
\end{align}

Furthermore, the one particle state is:

\begin{align}
 |l, \lambda\rangle = \sqrt{2 l^+} b_\lambda^\dagger(l) |0\rangle~,
\end{align}

with the normalisation:

\begin{align}
 \langle l, \lambda| l', \lambda'\rangle = 2(2 \pi)^3 l^+   
\delta(l^+-l'^+) 
\delta^{(2)}(\bf 
l_\perp-\bf l'_\perp )\delta_{\lambda, \lambda'}  ~. 
\end{align}

Moreover,  for  unpolarized dPDFs, 
 the following relations are usually the relevant ones:

\begin{align}
& \bar u_{\downarrow}(l) \gamma^+ u_{\downarrow}(k)=\bar 
u_{\uparrow}(l) 
\gamma^+ u_{\uparrow}(k) = 2 \sqrt{(l^+ k^+)}
\\
\nonumber
&\bar u_{\downarrow}(l) \gamma^+ u_{\uparrow}(k)=\bar u_{\uparrow}(l) 
\gamma^+ u_{\downarrow}(k) = 0,
\end{align}
where in the above equation $u_{\sigma}(k)=u_q(k,\sigma)$.
Now, for $\gamma^0$ one gets:

\begin{align}
\label{gamma0}
& \bar u_{\downarrow}(l) \gamma^0 u_{\downarrow}(k) =  \sqrt{(l^+ k^+)} 
+ \dfrac{(l_x+il_y) (k_x-ik_y)}{\sqrt{l^+ k^+} }
\\
\nonumber
& \bar u_{\uparrow}(l) \gamma^0 u_{\uparrow}(k) =  \sqrt{(l^+ k^+)} + 
\dfrac{(l_x-il_y) (k_x+ik_y)}{\sqrt{l^+ k^+} }
\\
\nonumber
&\bar u_{\downarrow}(l) \gamma^0 u_{\uparrow}(k)= - \bar 
u_{\uparrow}(l) 
\gamma^0 u_{\downarrow}(k) \sim 0~. 
\end{align}
The last relation is 0  since 
in the IMF the term $m/P^+$  can be neglected as well as 
for the 
$\gamma^+$ case.
The first line of the above equation is similar to that obtained within 
the 
light-cone treatment case, but the main difference is the presence 
of the 
transverse components of the quark momenta. However, by using momentum 
conservation, Eq. (\ref{gamma0}) can be written in terms of 
  $k^+ = P^+ x_1$ and $l^+ = \bar 
x_1 P^+$:

\begin{align}
  \sqrt{l^+ k^+} 
+ \dfrac{(l_x+il_y) (k_x-ik_y)}{\sqrt{l^+ k^+} } = P^+ \sqrt{x_1 
\bar x_1} + \dfrac{(l_x+il_y) (k_x-ik_y)}{ P^+ \sqrt{x_1\bar x_1} }~.
\end{align}
 Since, by using the standard LF procedure discussed in Sect. \ref{II}, 
the 
dPDF does not depends on the meson frame, the 
 $P^+$ dependence  is completely simplified. 
  Such a procedure leads to 
rewrite 
the 
correction 
due to the choice of $\gamma^0$ as follows:

\begin{align}
 \mathcal{O_\gamma} \propto  \dfrac{(l_x+il_y) (k_x-ik_y)}{ (P^+)^2 }~.
\end{align}
In the IMF such a contribution is suppressed. Let us remind 
that for a bound confined system, the w.f. goes to zero for $ 
{\bf l_\perp}~\mbox{and}~ {\bf k_\perp}$ very large.  The other 
source of 
 difference between the calculation performed within the lattice 
condition w.r.t. the standard light-cone case, comes from the choice of 
the  the quark field separation, i.e.  $y^0=0$, instead of $y^+=0$, see 
 Eqs. (\ref{f2c1}) and (\ref{lattice1}).
By working in the lattice conditions, one needs to evaluate:

\begin{align}
\nonumber
 \int dy_z~ e^{y (k_1-\bar k_1)} = e^{{\bf y}_\perp \cdot ({\bf k_{1 
\perp}-{\bf \bar k_{1 \perp}} })  } &~ \int dy^- e^{y^-(k_1^+-\bar 
k_1^+) 
} 
 e^{y^+(k_1^--\bar k_1^-) }
\\
=e^{{\bf y}_\perp \cdot ({\bf k_{1 
\perp}-{\bf \bar k_{1 \perp}} })  }&\int dy^-~ e^{y^-(k_1^+-\bar k_1^+ 
- k_1^- +\bar k_1^-) }~,
\label{deltan}
\end{align}
where we have used that for $y_0=0$ one gets $y_z = y^+ =-y^-$. 
We 
recall that $k$ and $\bar k$ are the momentum of partons in the hadron 
in the initial and final states, respectively. Thus from Eq. 
(\ref{deltan}), we get:

\begin{align}
 \delta(k_1^+-\bar k_1^+ - k_1^- +\bar k_1^-) = \dfrac{1}{P^+} 
\delta(x_1-\bar x_1-  \mathcal{O}_y )~,
\end{align}
where $\mathcal{O}_y$ represents the corrections due to the choice of 
$y_0=0$ w.r.t. $y^+=0$. One can show that this quantity reads:

\begin{align}
 \mathcal{O}_y = \dfrac{m^2+k_\perp^2}{2 (P^+)^2  x_1} - 
\dfrac{m^2+ \bar k_\perp^2}{2 (P^+)^2 \bar x_1}~.
\end{align}
In this case the correction is proportional to $1/(P^+)^2$. In the IMF 
such a contribution is small. 
Thus, from only a kinematic point of view,
the light-cone expression of dPDFs is similar to that obtained within 
the lattice framework in the IMF. In closing we stress that also in the 
analysis of Ref. \cite{lattice}, the authors claim that the standard 
expression (\ref{appmom}), which relates the eff to the product of 
one-body quantities,  
is restored for $q^2 <<m_\pi^2$.

\end{document}